\documentclass[a4paper,11pt,reqno]{amsart}

\usepackage{amsmath}
\usepackage{amssymb}
\usepackage{bm}
\usepackage{graphicx}
\usepackage[latin1]{inputenc}
\usepackage{amsfonts}
\usepackage{amsthm}
\usepackage{newlfont}
\usepackage[round,comma]{natbib}
\usepackage{latexsym}
\usepackage{float}
\usepackage{capt-of}
\usepackage{stackrel}
\usepackage{mathrsfs}

\newcommand{\ops}{\mathcal{S}}
\newcommand{\fof}{Fredholm of index 0}
\newcommand{\myalpha}{\zeta}
\newcommand{\mybeta}{\tau}
\newcommand{\mya}{\mathscr{Z}}
\newcommand{\myb}{\mathscr{T}}
\newcommand{\myH}{\mathcal{H}}
\newcommand{\myB}{\mathcal{L}_\mathcal{H}}

\DeclareMathOperator{\Ima}{Im}
\DeclareMathOperator{\Ker}{Ker}

\newcommand{\ar}{AR with a unit root of finite type}
\newcommand{\ars}{ARs with a unit root of finite type}

\newcommand{\as}{attractor space}
\newcommand{\cs}{cointegrating space}
\newcommand{\dd}{\mathrm{d}}

\numberwithin{equation}{section}

\newlength{\defbaselineskip}
\newcommand{\setlinespacing}[1]{\setlength{\baselineskip}{#1 \defbaselineskip}}

\newtheorem{thm}{Theorem}[section]
\newtheorem{coro}[thm]{Corollary}
\newtheorem{lemma}[thm]{Lemma}
\newtheorem{prop}[thm]{Proposition}
\newtheorem{ass}[thm]{Assumption}

\newtheorem{defn}[thm]{Definition}

\theoremstyle{remark} 
\newtheorem{remark}[thm]{Remark}
\theoremstyle{remark} 

\theoremstyle{definition}

\newcommand{\E}{\operatorname{E}}

\newcommand{\rank}{\operatorname{rank}}

\newcommand{\msp}{\operatorname{sp}}
\newcommand{\csp}{\overline{\operatorname{sp}}}

\newcommand{\tref}[1]{$\ref{#1}$}
\newcommand{\rimp}{\Rightarrow}

\newcommand{\miff}{\Leftrightarrow}
\newcommand{\cp}{\clearpage}
\newcommand{\BC}{\mathbb{C}}
\newcommand{\BR}{\mathbb{R}}

\newcommand{\BZ}{\mathbb{Z}}

\newcommand{\rf}{rank factorization}

\newcommand{\wt}[1]{\widetilde{#1}}

\newcommand{\pole}[1]{\textsc{pole}{$(#1)$}}

\usepackage[colorlinks,citecolor=blue,urlcolor=blue]{hyperref}

\begin{document}

\setlinespacing{1.4}

\title[]{Cointegration in functional autoregressive processes}%
\author[]{Massimo Franchi and Paolo Paruolo\\ \\ \today}
\thanks{\\ M. Franchi, Sapienza University of Rome, P.le A. Moro 5, 00185 Rome, Italy; e-mail: \texttt{massimo.franchi@uniroma1.it}. \\ \\P. Paruolo, European Commission, Joint Research Centre (JRC), Via E.Fermi 2749, I-3027 Ispra (VA), Italy; e-mail: \texttt{paolo.paruolo@ec.europa.eu}, \textbf{corresponding author}.\\
\\The paper benefited from useful comments from the Editor, Peter C.B. Phillips, three anonymous referees and participants at 2018 NBER-NSF Time Series Conference, University of California San Diego. The first author acknowledges partial financial support from MIUR PRIN grant 2010J3LZEN.\\
The idea of the present paper was conceived while the first author was visiting the Department of Economics, Indiana University, in January 2017; the hospitality of Yoosoon Chang and Joon Park is gratefully acknowledged. During the revision of the paper in September 2018, the first author visited the Department of Economics, University of California San Diego, and the hospitality of Brendan K. Beare is gratefully acknowledged.
}


\begingroup
\def\uppercasenonmath#1{} 
\maketitle
\endgroup

\begin{abstract}
This paper defines the class of $\myH$-valued autoregressive (AR) processes with a unit root of finite type, where $\myH$ is an infinite dimensional separable Hilbert space, and derives a generalization of the Granger-Johansen Representation Theorem valid for any integration order $d=1,2,\dots$. An existence theorem shows that the solution of an \ar\ is necessarily integrated of some finite integer $d$ and displays a common trends representation with a finite number of common stochastic trends of the type of (cumulated) bilateral random walks and an infinite dimensional \cs. A characterization theorem clarifies the connections between the structure of the AR operators and $(i)$ the order of integration, $(ii)$ the structure of the \as\ and the \cs, $(iii)$ the expression of the cointegrating relations, and $(iv)$ the Triangular representation of the process. Except for the fact that the number of cointegrating relations that are integrated of order 0 is infinite, the representation of $\myH$-valued \ars\ coincides with that of usual finite dimensional VARs, which corresponds to the special case $\myH=\BR^p$.
\end{abstract}

\section{Introduction}\label{sec_introduction}

The theory of time series that take values in infinite dimensional separable Hilbert spaces, or $\myH$-valued processes, is 
receiving increasing attention in econometrics.
$\myH$-valued processes allow to represent directly the dynamics of infinite-dimensional objects, such as Lebesgue square-integrable functions on a compact domain. In this way, they allow greater modeling generality with respect to models for conditional means and variances, see e.g. \citet{HK:12}.

One notable special case is given by $\myH$-valued processes $h=\psi (f)$, where $f$ a generic probability density function (pdf) and $\psi$ is an invertible transformation; the transformation is needed because the space of pdfs is convex but not linear, see \cite{petersen2016}. Modeling dynamics of a whole pdf
appears of interest e.g. for the income distribution, see e.g. \citet{WB:05}, \citet{Piketty:14} and \citet{CKP:16}.

An important early contribution to the theory of functional time series is \citet{Bos:00}, where a theoretical treatment of linear processes in Banach and Hilbert spaces is developed. There, emphasis is given to the derivations of laws of large numbers and central limit theorems that allow to discuss estimation and inference for $\myH$-valued stationary autoregressive (AR) models.

Economic applications of functional time series analysis include studies on the term structure of interest rates, see \citet{KO:08}, and intraday volatility, see \citet{HHR:13} and \citet{GHK:13}; additional applications can be found in the recent monographs \citet{HK:12} and \citet{KR:17} and in the review article \citet{HK:12b}.

Recently \citet{CKP:16} applied Functional Principal Components Analysis (FPCA) directly on the space of densities for individual earnings and intra-month distributions of stock returns.\footnote{\citet{Bea:17} pointed out the issue that the space of density is not linear, see also \citet{BS:18b}.} They found evidence of unit root persistence in a handful of coordinates of these cross-sectional distributions. The framework proposed by \citet{CKP:16} has (by construction) a finite number of $I(1)$ stochastic trends and an infinite dimensional \cs. The theory is developed starting from the infinite moving average representation of the first differences of the process and the potential unit roots are identified and tested through FPCA.

Representation of $\myH$-valued AR processes with unit roots has been recently considered in the literature. \citet{HP:WP} consider $\myH$-valued AR$(1)$ processes with compact operator and prove that an extension of the Granger-Johansen Representation Theorem, see Theorem 4.2 in \citet{Joh:96}, holds in the $I(1)$ case. The corresponding common trends representation, or functional Beveridge-Nelson decomposition, displays a finite number of $I(1)$ stochastic trends and an infinite dimensional \cs. They further propose an estimator for the functional autoregressive operator which builds on the results in \citet{CKP:16}.

\citet{BSS:17} consider $\myH$-valued AR$(k)$, $k\geq 1$, with compact operators if $k>1$ and no compactness assumption if $k=1$, and show that the Granger-Johansen Representation Theorem holds in the $I(1)$ case. If $k>1$, the number of $I(1)$ stochastic trends is finite and the dimension of the \cs\ is infinite, while if $k=1$ this is not necessarily the case. In order to obtain the common trends representation of $\myH$-valued AR$(k)$, $k\geq 1$, with compact operators.
\citet{BS:18} are the first to employ a theorem on the inversion of analytic operator functions in \citet{GGK:90:1}.\footnote{The same theorem is used here to discuss the existence of a common trends representation in Section \tref{sec_ars}.} They 
also present results on the $I(2)$ case that show that the number of $I(2)$ stochastic trends is finite and the dimension of the \cs\ is infinite.

Finally, \citet{CHP:WP} consider an error correction form with compact error correction operator and show that in this case the number of $I(1)$ stochastic trends is infinite and the dimension of the \cs\ is finite. Moreover, they show that Granger-Johansen Representation Theorem continues to hold.

This paper considers a more general class of AR processes, called the class of \ars. This class contains $\myH$-valued ARs with compact operators as a special case. This paper derives a generalization of the Granger-Johansen Representation Theorem for this class, valid for any integration order $d=1,2,\dots$ .

An existence theorem is provided; this shows that the solution of an \ar\ is necessarily $I(d)$ for some finite integer $d$ and displays a common trends representation with a finite number of common stochastic trends of the type of (cumulated) bilateral random walks and an infinite dimensional \cs. This result is a direct consequence of a well known theorem in operator theory,
and first employed in \citet{BS:18} in the context of $\myH$-valued ARs with compact operators.

Despite these interesting implications, this existence result does not address a number of important issues, such as the connections between the structure of the AR operators and $(i)$ the order of integration of the process, $(ii)$ the structure of the \as\ and the \cs\ and $(iii)$ the expression of the cointegrating relations.
The characterization of these links in the generic $I(d)$ case constitutes the main contribution of the present paper. More specifically, a necessary and sufficient condition for the order of integration $d$ is given in terms of the decomposition of the space $\myH$ into the direct sum of $d+1$ orthogonal subspaces $\mybeta_h$, $h=0,\dots,d$, that are expressed recursively in terms of the AR operators. This condition is called the `\pole{d} condition', because it is a necessary and sufficient condition for the inverse of the $A(z)$ function to have a pole of order $d$ at $z=1$.

A crucial feature of the present \pole{d} conditions is that the subspaces in the orthogonal direct sum decomposition $\myH=\mybeta_0 \oplus \mybeta_1 \oplus \cdots \oplus \mybeta_d$, $\mybeta_d\neq \{ 0 \}$, identify the directions in which the properties of the process differ. Specifically, for any nonzero $v\in \mybeta_0$, which is infinite dimensional, one can combine $\langle v,x_t\rangle$ with differences 
$\Delta^n x_{t}$ for $n=1,\dots,d-1$ to find $I(0)$ polynomial cointegrating relations. For any nonzero $v\in \mybeta_1$, with dimension $0 \leq \dim \mybeta_1 < \infty $, one can combine $\langle v,x_t\rangle$ 
with differences 
$\Delta^n x_{t}$ for $n=1,\dots,d-2$ and find $I(1)$ polynomial cointegrating relations.

This kind of feature is valid for $\mybeta_{2}, \dots, \mybeta_{d-2}$; for $\mybeta_{d-1}$, with $0 \leq \dim \mybeta_{d-1} < \infty $, one has $\langle v,x_t\rangle \sim I(d-1)$ for any nonzero $v$ in $\myH$, without
polynomial cointegration. Finally for nonzero $v$ in $\mybeta_{d}$, with
with $0 < \dim \mybeta_{d} < \infty $
one has $\langle v,x_t\rangle \sim I(d)$ for any nonzero $v$, i.e. all $v$-characteristics have no cointegration.
These results parallel the ones in the Triangular Representation in the finite dimensional case $\myH=\BR^p$ discussed in \citet{Phi:91} and \citet{SW:93}; see also \citet{FP_idcoeff}.

These results show that conditions and properties of \ars\ extend those that apply in the usual finite dimensional VAR case; in particular for $\myH=\BR^p$ one finds the $I(1)$ and $I(2)$ results in \citet{Joh:96}, and for the generic $I(d)$ case, one finds the results in \citet{FP_idcoeff}. Except for the fact that the number of $I(0)$ cointegrating relations is infinite,  the infinite dimensionality of $\myH$ does not introduce additional elements in the representation analysis of \ars.

The present results are based on orthogonal decomposition of the embedding Hilbert space. Orthogonal and non-orthogonal projections are well known concepts in econometrics. Students are usually introduced to these concepts when learning OLS and GLS, where the choice between the two is usually discussed in terms of estimation efficiency; see \citet{Phi:91b} for how these arguments are modified for spectral GLS regressions methods in a cointegration context.
In the context of the representation theory considered here, results can be obtained using either orthogonal or non-orthogonal projections. The present choice of orthogonal projections is found to ease exposition and to simplify the characterization of the cointegrating $v$-characteristics of the process.

The rest of the paper is organized as follows: Section~\tref{sec_setup} presents basic definitions and concepts, Section~\tref{sec_ars} discussed the assumption of unit root of finite type and reports initial existence results for a pole of finite order;
Section~\tref{sec_char_1_2} provides a characterization of $I(1)$ and $I(2)$ \ars\ and Section~\tref{sec_char_d} extends the analysis
to the general $I(d)$, $d=1,2,\dots$, case. Section~\tref{sec_CONC} concludes.

Three Appendices collect background definitions, novel inversion results and proofs of the statements in the paper. Specifically,
Appendix \tref{app_dot} reviews notions on operators acting on a separable Hilbert space $\myH$ and on $\myH$-valued random variables; Appendix \tref{app_lemmas} presents novel results on the inversion of a meromorphic operator function and 
Appendix \tref{app_text} reports proofs of the results in the paper.

\section{$\myH$-valued linear process, order of integration and cointegration}\label{sec_setup}

This section introduces the notions of weakly stationary, white noise, linear, integrated, and cointegrated processes that take values in a separable Hilbert space $\myH$, where separable means that $\myH$ admits a countable orthonormal basis. Basic definitions of operators acting on $\myH$ and of $\myH$-valued random variables are reported in Appendix \tref{app_dot}.

\subsection{Definitions}
The definitions of weakly stationary and white noise process are taken from \citet[][Definitions 2.4, 3.1, 7.1]{Bos:00}, while those of linear, integrated and cointegrated process are adapted from \citet{Joh:96}; they are similar to those employed in \citet{CKP:16,BSS:17,BS:18}. The definition of expectation $\E(\cdot)$, covariance operator and cross-covariance function used in the following are reported in Appendix \tref{app_random}.

\begin{defn}[Weakly stationary process]\label{def_WS}
An $\myH$-valued stochastic process $\{\varepsilon_t,t\in \BZ\}$ is said to be \emph{weakly stationary} if $(i)$ $0<E(\|\varepsilon_{t}\|^2)<\infty$, $(ii)$ $\E(\varepsilon_t)$ and the covariance operator of $\varepsilon_t$ do not depend on $t$ and $(iii)$ the cross-covariance function of $\varepsilon_t$ and $\varepsilon_s$, $c_{\varepsilon_t,\varepsilon_s}(h,v)$, is such that $c_{\varepsilon_t,\varepsilon_s}(h,v)=c_{\varepsilon_{t+u},\varepsilon_{s+u}}(h,v)$ for all $h,v\in \myH$ and all $s,t,u\in \BZ$.
\end{defn}

The notion of $\myH$-valued white noise is introduced next.

\begin{defn}[White noise process]\label{def_WN}
An $\myH$-valued weakly stationary stochastic process $\{\varepsilon_t,t\in \BZ\}$ is said to be \emph{white noise} if $(i)$ $\E(\varepsilon_t)=0$ and $(ii)$ $c_{\varepsilon_t,\varepsilon_s}(h,v)=0$ for all $h,v\in \myH$ and all $s\neq t,s,t\in \BZ$, where $c_{\varepsilon_t,\varepsilon_s}(h,v)$ is the cross-covariance function of $\varepsilon_t$ and $\varepsilon_s$; it is called \emph{strong white noise} if $(i)$ holds, and $(ii)$ is replaced by the requirement that $\varepsilon_t$ is an i.i.d. sequence of $\myH$-valued random variables.
\end{defn}

Note that by definition any strong white noise is white noise, and any white noise process is weakly stationary. The same property holds for linear combinations of lags of a white noise process with suitable weights; this leads to the class of linear processes, introduced in Definition \tref{def_LP} below.

In the definition below, the following notation is employed: $D(z_0,\rho)$ denotes the open disc $\{z\in \BC:|z-z_0|<\rho\}$ with center $z_0\in\BC$ and radius $0<\rho\in \BR$ and $\myB$ indicates the set of bounded linear operators on $\myH$ with norm $\| A \|_{\myB}=\sup_{\| v \|=1}\| Av \|$; an operator function $B(z)=\sum_{n=0}^{\infty}B_{n}(z-z_0)^n$, where $B_{n} \in \myB$, is said to be absolutely convergent on $D(z_0,\rho)$ if $\sum_{n=0}^{\infty}\|B_{n}\|_{\myB}|z-z_0|^n<\infty$ for all $z\in D(z_0,\rho)$.\footnote{Note that $\sum_{n=0}^{\infty}\|B_{n}\|_{\myB}|z-z_0|^n<\infty$ for all $z\in D(z_0,\rho)$ implies that $\sum_{n=0}^{\infty}B_{n}(z-z_0)^n$ converges in the operator norm to $B(z)\in\myB$ for all $z\in D(z_0,\rho)$, i.e. $\|\sum_{n=0}^{N}B_{n}(z-z_0)^n-B(z) \|_{\myB}\rightarrow 0$ as $N\rightarrow \infty$.} The lag operator is denoted by $L$ and $\Delta = 1-L$ is the difference operator.

\begin{defn}[Linear process]\label{def_LP}
Let $\{\varepsilon_{t},t\in \BZ\}$ be white noise; an $\myH$-valued stochastic process $\{u_t,t\in \BZ\}$ with expectation $\{\mu_{t},t\in \BZ\}$, $\mu_{t}=\E(u_{t})$, is said to be a linear process if
$$
u_t-\mu_{t}=\sum_{n=0}^{\infty}B_{n}\varepsilon_{t-n},\qquad B_n\in\myB,\qquad B_0=I,
$$
where $B(z)=\sum_{n=0}^{\infty}B_{n}z^n$, $z\in \BC$, is absolutely convergent on the open disc $D(0,\rho)$ for some $\rho>1$.
\end{defn}

As discussed in Section 7.1 in \citet{Bos:00}, existence and weak stationarity of $u_t-\mu_{t}=\sum_{n=0}^{\infty}B_{n}\varepsilon_{t-n}$ are guaranteed by the square summability condition $\sum_{n=0}^{\infty}\|B_{n}\|_{\myB}^2<\infty$. Observe that the requirement that $B(z)$ is absolutely convergent on $D(0,\rho)$ for some $\rho>1$ is stronger. In fact, $\sum_{n=0}^{\infty}\|B_{n}\|_{\myB}|z|^n<\infty$ for all $z\in D(0,\rho)$, $\rho>1$, implies $\sum_{n=0}^{\infty}\|B_{n}\|_{\myB}<\infty$ and hence $\sum_{n=0}^{\infty}\|B_{n}\|_{\myB}^2<\infty$. This shows that $u_t-\mu_{t}$ in Definition \tref{def_LP} is well defined and weakly stationary.

Moreover, $B(z)\in \myB$ for all $z\in D(0,\rho)$, $\rho>1$, implies that $B(1)$ is a bounded linear operator. Finally note that $B(z)$ is infinitely differentiable on $D(0,\rho)$, $\rho>1$, and the series obtained by termwise $k$ times differentiation, $\sum_{n=k}^{\infty}n(n-1)\cdots (n-k+1)B_{n}z^{n-k}$, is absolutely convergent and coincides with the $k$-th derivative of $B(z)$ for each $z\in D(0,\rho)$. Hence $\sum_{n=k}^{\infty}n(n-1)\cdots (n-k+1)\|B_{n}\|_{\myB}<\infty$, which for $k=1$ reads $\sum_{n=1}^{\infty}n\|B_{n}\|_{\myB}<\infty$; this condition is employed in \citet{CKP:16}.

The notions of integration and integral operator are introduced next. 

\begin{defn}[Order of integration]\label{def_INT}
A linear process $u_t-\mu_{t}=B(L)\varepsilon_{t}$ is said to be integrated of order $0$, written $u_t\sim I(0)$, if $B(1)\neq 0$. If $\Delta^d z_t$ is $I(0)$ for some finite integer $d=1,2,\dots$, $\{z_{t},t\in \BZ\}$ is said to be integrated of order $d$, indicated $z_t \sim I(d)$.
\end{defn}

This definition coincides with Definition 3.3 in \citet{Joh:96} of an $I(d)$ process for the special case $\myH=\BR^p$.

Observe that a white noise process is $I(0)$ and that an $I(0)$ process is weakly stationary. In order to see that a weakly stationary is not necessarily $I(0)$, take for instance $u_t=\varepsilon_{t}-\varepsilon_{t-1}$; this process is weakly stationary, with $B(1)=0$ and hence it does not satisfy the definition of an $I(0)$ process, showing that the two concepts do not coincide. The distinction between weak stationarity and $I(0)$-ness is relevant for the definition of order of integration: in fact, the cumulation of an $I(0)$ process is necessarily $I(1)$ while the cumulation of stationary process is not necessarily so.

Following \citet{HP:WP}, one can define the $v$-characteristic of $x_t$ as the scalar process $\langle v, x_t \rangle$, for any $v \in \myH$. From Definition \tref{def_INT}, one can see that a generic $v$-characteristic of $x_t \sim I(d)$ is itself at most integrated of order $d$; the case when a $v$-characteristic of $x_t \sim I(d)$ is integrated of lower order $b<d$ is associated with the notion of cointegration.

\begin{defn}[Cointegrated process]\label{def_COINT}
An $I(d)$ process $z_t$ is said to be cointegrated if there exists a nonzero $v$-characteristic
$v\in \myH$ such that $\langle v,z_t\rangle$ is $I(b)$ for some $b<d$. The set $\{v\in \myH:\langle v,z_t \rangle\sim I(b),b<d\}\cup \{0\}$ is called the \cs\ and its orthogonal complement is called the \as.
\end{defn}

As in the usual finite dimensional case, $z_t$ is cointegrated if there exists a nonzero linear combination $v$ of $z_t$ (i.e. a $v$-characteristic of $z_t$) that has lower order of integration than the original process. Observe that the \as\ (respectively the \cs) contains $0\in \myH$ and all nonzero $v\in \myH$ that correspond to a $v$-characteristic of $z_t$ with the same (respectively lower) order of integration of the original process $z_t$. 
The null vector $0\in \myH$ is added so as to make the \cs\ a vector spaces.

The cases that have been studied in the literature correspond to finite dimensions either for the attractor or for the \cs. When both of them have finite dimension, $\myH$ is finite dimensional, so that the standard results in the literature apply. The case in which the \as\ is infinite dimensional and the \cs\ is finite dimensional corresponds to a process with an infinite number of $I(d)$ stochastic trends and a finite dimensional \cs. For $d=1$, this case has been discussed in \citet{CHP:WP} and in \citet{BSS:17} for $k=1$, see Proposition \tref{prop_BSS_k_1} below.

Most of the contributions in the literature have studied instead the case of an \as\ of finite dimension and a \cs\ of infinite dimension, i.e. the case where the process has a finite number of $I(d)$ $v$-characteristics and an infinite number of $I(b)$ $v$-characteristics with $b<d$. This is the setup studied in \citet{CKP:16}, \citet{HP:WP} and \citet{BSS:17} for $d=1$ and in \citet{BS:18} for $d=1$ and $d=2$. This is the setting considered in the present paper as well, and it is motivated also by the next example.

\subsection{Yield curve example}\label{sec_yield_curve}
As an example of a Hilbert space of economic interest, consider the yield curve $x_{\circ,t}(s)$, where $s$ denotes maturity and $t$ time. In this section $t$ is omitted, unless needed for clarity.

Let $\myH$ be the set of Lebesgue measurable functions $x_{\circ}(s)$ such that $\int_{0}^{s_{\max}}x_{\circ}^2(s)\dd s<\infty $, where $s_{\max}$ is the maximal maturity. 
One can rescale the maturity $s$ into $u=s/s_{\max}$ and define the rescaled yield curve $x(u)$ by $x(u)=x_{\circ}(u \cdot s_{\max})$,
with $u\in (0,1]$ and $\int_{0}^{1}x^{2}(u)\dd u<\infty $. The vector space operations on $\myH$ are defined in a natural way as $(x+y) (u) =x(u) +y(u) $ and $\left( \alpha x\right) (u) =\alpha x(u) $ where $\alpha \in \BR$. Next define the inner product
\begin{equation}\label{eq_inner_int}
\langle x,y\rangle=\int_{0}^{1}x(u)y(u)\dd u.
\end{equation}
This space of Lebesgue square-integrable functions equipped with the inner product \eqref{eq_inner_int} is a complete, separable Hilbert space, see e.g. \citet[][p. 214]{KR:17}.

The yield curve is often described in terms of the three features of level, slope and curvature, see e.g. \citet{CP:05}. These features of the yield curve can be associated with the following $v$-characteristics of $x$. Define $\pi _{j,1},\dots ,\pi _{j,j}$ as a partition of the unit interval $(0,1]$ into $j$ segments $\pi _{j,i}$ of length $1/j$, $\pi _{j,i}=(\frac{i-1}{j},\frac{i}{j}]$, and let $1_{\{u\in \pi _{j,i}\}}$ be the indicator function that takes value one when $u\in \pi _{j,i}$ and equals 0 otherwise.

Next define the following $v$ functions
\begin{align*}
v_{0} &=1_{\{u\in \pi _{1,1}\}}, \qquad v_{1} =\frac{1}{2}\left( 1_{\{u\in \pi _{2,2}\}}-1_{\{u\in \pi_{2,1}\}}\right), & \\
v_{2} &=\frac{1}{4}\left( 1_{\{u\in \pi _{4,4}\}}-1_{\{u\in \pi_{4,3}\}}\right) - \frac{1}{4}\left( 1_{\{u\in \pi _{4,2}\}}-1_{\{u\in \pi_{4,1}\}}\right), &
\end{align*}
and observe that they belong to $\myH$, because they are Lebesgue square-integrable functions. Finally let $x$ denote the rescaled yield curve and note that
\begin{align*}
\langle v_{0},x\rangle &=\int_{0}^{1}v_{0}(u) x(u)\dd u=\int_{0}^{1}x(u)\dd u, \\
\langle v_{1},x\rangle &=\int_{0}^{1}v_{1}(u) x(u)\dd u=\frac{1}{2}\left(\int_{\frac{1}{2}}^{1}x(u)\dd u-\int_{0}^{\frac{1}{2}}x(u)\dd u\right), \\
\langle v_{2},x\rangle &=\int_{0}^{1}v_{2}(u) x(u)\dd u=\frac{1}{4}\left( \int_{\frac{3}{4}}^{1}x(u)\dd u-\int_{\frac{1}{2}}^{\frac{3}{4}}x(u)\dd u\right) -\frac{1}{4}\left( \int_{\frac{1}{4}}^{\frac{1}{2}}x(u)\dd u-\int_{0}^{\frac{1}{4}}x(u)\dd u\right).
\end{align*}

One can see that $\langle v_{0},x\rangle$ computes the average yield curve, and hence can be associated with the level of the yield curve. Similarly $\langle v_{1},x\rangle$ computes the difference between the average yield on the longer maturities and the one on the shorter maturities;\ hence it can be associated with the slope of the yield curve. Finally $\langle v_{2},x\rangle$ computes the difference of the slopes on the longer maturities and the shorter maturities; hence it can be associated with the curvature of the yield curve.

This shows that $v_{0}$, $v_{1}$, $v_{2}$ define interesting $v$-characteristics for the yield curve $x$. If the yield curve $x$ is modeled as a functional time series, $x_{t}$, then it is interesting to ask questions of the type: ``what is the order of integration of the level (or slope, or curvature) of the yield curve?''. These questions translate into ``what is the order of integration of the $v_{j}$-characteristics, $j=0,1,2$, of the yield curve $x_{t}$?''.

This illustrates how interesting hypotheses can be formulated in this context; clearly other types of hypotheses can be formulated in a similar way. Moreover, it is of interest to determine how many and which characteristics are nonstationary, which corresponds to estimating the (dimension of the) \as.

It appears natural in this context to assume (or test) that there are only a finite number of factors driving the dynamics of the yield curve. This translated into the hypothesis that there are only a finite number of nonstationary $v$-characteristics; in this case, $x_t$ would have a finite dimensional \as\ and an infinite dimensional \cs. This seems to be a reasonable assumption to be tested empirically also beyond the case of the yield curve; this case is the one studied in the present paper, see Corollary \tref{coro_EXI_CI} below.


\section{\ars}\label{sec_ars}

This section introduces the class of $\myH$-valued ARs that is studied in the present paper, called \ars. It also presents an existence result about their common trends representation, which shows that the solution of an \ar\ is necessarily $I(d)$ for some finite integer $d$ and displays a common trends representation with a finite number of common stochastic trends of the type of (cumulated) bilateral random walks.\footnote{This result is a direct consequence of a well known theorem in operator theory, reported in Theorem \tref{thm_FLSF} in Appendix \tref{app_operator}, and first employed in \citet{BS:18} in the context of $\myH$-valued ARs with compact operators.}
The relations of \ars\ with the ARs studied in literature are also discussed in this section, and an example of an \ar\ with a non-compact AR operator is given.

\subsection{Main assumption}
Consider an $\myH$-valued AR process
\begin{equation}\label{eq_AR}
x_{t}=A^\circ_1 x_{t-1}+\dots+ A^\circ_k x_{t-k}+\varepsilon_{t},\qquad A^\circ_n\in\myB,\qquad t\in \BZ,
\end{equation}
where $I$ indicates the identity operator in $\myB$, $\{\varepsilon_{t},t\in \BZ\}$ is white noise and the operator function
$$
A(z)=I-\sum_{h=1}^kA^\circ_h z^h,\qquad z \in \BC,\qquad A(1)\neq 0,
$$
is non-invertible at $z=1$ and invertible in the punctured disc $D(0,\rho)\setminus\{1\}$ for some $\rho>1$.

This requirement restricts attention to unit roots at frequency zero, corresponding to the point $z=1$ on the unit disc. Note that there is no loss of generality in assuming that $A(1)\neq 0$. In fact, if $A(1)=0$, one can factorize
$(1-z)^s$ from $A(z)$,
$A(z)=(1-z)^s\wt{A}(z)$ for some $\wt{A}(1)\neq \{ 0 \}$ and some $s>0$, and rewrite the AR equations $A(L)x_{t}=\varepsilon_{t}$ as $\wt{A}(L)y_{t}=\varepsilon_{t}$ for $y_{t}=\Delta^s x_{t}$.

In order to state the key assumption, it is useful to expand the operator function $A(z)=I-\sum_{h=1}^k A^\circ_h z^h$ around 1, obtaining
\begin{equation}\label{eq_def_Az}
A(z)=\sum_{n=0}^{\infty}A_{n}(1-z)^n,\qquad
A_n =
\left\{
\begin{array}{cl}
I-\sum_{h=1}^k A^\circ_h & \mbox{ for } n=0 \\
(-1)^{n+1}\sum_{h=0}^{k-n}\binom{n+h}{n}A^\circ_{n+h} & \mbox{ for } n=1,2,\dots
\end{array}
\right.,
\end{equation}
where empty sums are defined to be 0 and hence $A_n = 0 $ for $n>k$.

The notion of eigenvalue of finite type, see \citet[][section XI.9]{GGK:90:1}, is central in the present setup and it is reported next. For any $A \in\myB$ the subspace $\{v\in \myH:Av=0 \}$, written $\Ker A$, is called the kernel of $A$ and the subspace $\{Av : v\in \myH\}$, written $\Ima A$, is called the image of $A$. The dimension of $\Ima A$, written $\dim \Ima A$, is called the rank of $A$.

\begin{defn}[Eigenvalue of finite type]\label{def_eig_finite_type}
A point $z_0\in \BC$ is said to be an eigenvalue of finite type of $A(z)$ if

\noindent$(i)$ $A(z_0)$ is Fredholm, i.e. $n=\dim \Ker A(z_0)<\infty$ and $q=\dim (\Ima A(z_0))^\bot<\infty$, of index $n-q$,

\noindent$(ii)$ $A(z_0)v=0$ for some nonzero $v\in \myH$,

\noindent$(iii)$ $A(z)$ is invertible for all $z$ in some punctured disc $D(z_0,\delta)\setminus\{z_0\}$.
\end{defn}

Direct consequences of this definition are listed in the following remark.

\begin{remark}\label{rem_eigFT0}
If $A(z)$ has an eigenvalue of finite type at $z=z_0$, $A(z_0)$ is necessarily \fof, see \citet[][Section XI.9]{GGK:90:1}. 
Combining this with $(i)$ and $(ii)$ in Definition \tref{def_eig_finite_type} one thus has that $0<\dim \Ker A(z_0)=\dim (\Ima A(z_0))^\bot<\infty$.\footnote{Remark that when $\myH$ is finite dimensional any operator is \fof\ and any eigenvalue is of finite type.} Moreover, $\Ima (A(z_0))$ is necessarily closed, see Theorem 2.1 in \citet[][Section 15.2]{GGK:03}, and hence, see Theorem 3 in \citet[][Chapter 9]{BI:03}, the generalized
maximal Tseng inverse of $A(z_0)$ exists, written $A(z_0)^+$, and it is unique. In the following
the `generalized maximal Tseng inverse' is abbreviated in the `generalized inverse'.
\end{remark}

The key assumption is introduced next.
\begin{ass}[\ar\ at $z=1$]\label{ass_finite_type}
Let $A(z)$ be as in \eqref{eq_AR} and \eqref{eq_def_Az}, with an eigenvalue of finite type at $z=1$ in the disc $z\in D(0,\rho)$, $\rho>1$; then $A(z)$ is said to be an \ar\ at $z=1$, or simply an \ar.
\end{ass}
That is, an \ar\ is such that $A(z)$ is invertible for all $z\in D(0,\rho)\setminus\{1\}$ for some $\rho>1$, $0<\dim \Ker A_0=\dim (\Ima A_0)^\bot<\infty$ and $\Ima A_0$ is closed, where $A_0$ is as in \eqref{eq_def_Az}.

\subsection{Existence of a common trends representation}\label{sec_exi}

Under Assumption \tref{ass_finite_type}, one can apply the results in Section XI.9 of \citet{GGK:90:1}, reported in Theorem \tref{thm_FLSF} in Appendix \tref{app_operator}, and first employed in \citet{BS:18} in the context of $\myH$-valued ARs with compact operators. These results guarantee that there exist a finite integer $d=1,2,\dots$ and finite rank operators $C_{0},C_{1},\dots,C_{d-1}$ such that
\begin{equation}\label{eq_LAU2}
A(z)^{-1}=\sum_{n=0}^{\infty} C_n (1-z)^{n-d}, \qquad z\in D(0,\rho)\setminus\{1\},\qquad \rho>1,
\end{equation}
so that the inverse of $A(z)$ has a pole of finite order $d$ at $z=1$.

This implies that the solution of the AR equations is $I(d)$ for some finite integer $d$. Moreover, because the operators that make up the principal part of $A(z)^{-1}$ around $z=1$ have finite rank, $x_t$ displays a common trends representation with a finite number of common stochastic trends of the type of (cumulated) bilateral random walks, as reported in Theorem \tref{thm_EXI} below.

In order to state Theorem \tref{thm_EXI}, the cumulation operator $\ops$ in introduced, following \citet{Gre:99}.
\begin{defn}[Integral operator $\ops $]\label{def_S}
For a generic process $\{w_{t},t\in \BZ\}$ the integral operator $\ops $ is defined as
\begin{equation}\label{eq_S_def}
\ops w_{t}=1_{(t\geq 1)}\cdot \sum_{i=1}^{t}w_{i}-1_{(t\leq -1)}\cdot \sum_{i=t+1}^{0}w_{i}.
\end{equation}
When $w_t=\varepsilon_t$ is white noise, the notation $s_{h,t}=\ops ^h \varepsilon_{t}$, $h=1,2,\dots$, is employed.
\end{defn}

Remark that by definition $\ops $ assigns value 0 to the cumulated process at time 0. In fact, applying the definition, also see Properties 2.1, 2.2 in \citet{Gre:99}, one has
\begin{equation}\label{eq_D_S}
\Delta \ops w_{t}=w_{t},\qquad \ops \Delta w_{t}=w_{t}-w_{0},\qquad t\in\BZ.
\end{equation}
Eq. \eqref{eq_D_S} shows that $\ops $ applied to $\Delta w_{t}$ regenerates the level of the process $w_{t}$, up to a constant; this parallels the constant of integration in indefinite integrals. The integral operator $\ops $ is hence the inverse of the difference operator $\Delta$ up a constant, which is set by Definition \tref{def_S} so as to make the cumulated process $\ops \Delta w_{t}$ equal to 0 at time 0.

Note that when $w_t=\varepsilon_t$ is white noise, \eqref{eq_S_def} implies that $s_{1,t}=\ops \varepsilon_t$ is a bilateral $\myH$-valued random walk, see \citet[][example 1.9 on p. 20]{Bos:00}; because $\Delta s_{1,t}=\Delta \ops \varepsilon_t=\varepsilon_{t}$ is $I(0)$, this shows that $s_{1,t}$ is $I(1)$. Similarly, for $h=2,3,\dots$, $s_{h,t}=\ops s_{h-1,t}\sim I(h)$ is the $(h-1)$-fold cumulation of the bilateral random walk $s_{1,t}\sim I(1)$. 

The following results connects \ars with the existence of a common trend representation in terms of stochastic trends of the above type.

\begin{thm}[Existence of a common trends representation]\label{thm_EXI}
Let $A(L)x_{t}=\varepsilon_{t}$ be an \ar. Then there exist a finite integer $d=1,2,\dots$ and finite rank operators $C_{0},C_{1},\dots,C_{d-1}$ such that $x_t$ has common trends representation
\begin{equation}\label{eq_CT}
x_t=C_{0}s_{d,t}+C_{1}s_{d-1,t}+\dots+C_{d-1}s_{1,t}+y_{t}+\mu_t,\qquad t \in \BZ,
\end{equation}
where $s_{h,t}=\ops ^h \varepsilon_{t}\sim I(h)$ is the $(h-1)$-fold cumulation of the bilateral random walk $s_{1,t}\sim I(1)$, $y_{t}=C_d^{\star}(L)\varepsilon_{t}$ is a linear process, $\mu_t=\sum_{n=0}^{d-1}v_{n}t^n$ is the expectation of $x_t$,
where $v_0,\dots,v_{d-1}\in\myH$ depend on the initial values of $x_{t},y_{t},\varepsilon_{t}$ for $t=-d,\dots,0$.
\end{thm}

In the common trends representation \eqref{eq_CT} the operators $C_{0},C_{1},\dots,C_{d-1}$ have finite rank; this implies that $x_t$ depends only on a finite number of bilateral (cumulated) random walks. In fact, these common stochastic trends are selected from $s_{h,t} \sim I(h)$, $h=1,\dots,d$, by the finite rank operators $C_{0},C_{1},\dots,C_{d-1}$ that load onto $x_t$ only a finite number of characteristics from $s_{h,t}$, $h=1,\dots,d$.

Theorem \tref{thm_EXI} implies a number of properties for \ars, some of which are listed in the following corollary.

\begin{coro}[Cointegration properties]\label{coro_EXI_CI}
Let $A(L)x_{t}=\varepsilon_{t}$ be an \ar. Then

\noindent$(i)$ $x_{t}\sim I(d)$ for some finite integer $d=1,2,\dots$,

\noindent$(ii)$ $x_t\sim I(d)$ is cointegrated,

\noindent$(iii)$ $\Ima C_0$ is the finite dimensional \as,

\noindent$(iv)$ $(\Ima C_0)^\bot$ is the infinite dimensional \cs.
\end{coro}
Corollary \ref{coro_EXI_CI} lists some implications of Theorem \tref{thm_EXI}, namely that $d$ (the order of the pole of the inverse of $A(z)$ at $z=1$) is finite, the process is cointegrated, the number of common trends is finite and the number of cointegrating relations is infinite.

Despite these interesting implications of Theorem \tref{thm_EXI}, these existence results do not address a number of important issues, such as
the connection between the structure of $A(z)$ and the order of integration $d$ of the process.
In fact, one cannot determine the order of integration of the solution of the AR equations using Theorem \tref{thm_EXI}. Moreover, Theorem \tref{thm_EXI} does not specify the connection between $\Ima C_0$ and the AR operators, so that one does not know how to construct the \as\ and the \cs\ in terms of the AR operators. Finally, the relations among the finite rank operators $C_{0},C_{1},\dots,C_{d-1}$ are not specified and hence Theorem \tref{thm_EXI} is silent about the structure of the cointegrating relations.

These additional characterization results form the main contribution of the present paper; they go beyond Theorem \tref{thm_EXI} and Corollary \tref{coro_EXI_CI}, and they are presented in full generality
in Section \tref{sec_char_d} for the generic $I(d)$ case. For ease of presentation, Section \tref{sec_char_1_2} starts with the $I(1)$ and $I(2)$ cases.

\subsection{Relations with the literature}

Before turning to these results, the present section discusses the relationship between Assumption \tref{ass_finite_type} and the assumptions employed in the literature. An example in the next section illustrates the differences.

The following proposition discusses the relation with \citet{CKP:16}, who study $I(1)$ processes $x_t$ satisfying $\Delta x_t=B(L)\varepsilon_{t}$, where $\sum_{n=1}^{\infty}n\|B_{n}\|_{\myB}<\infty$ and $\dim \Ima B(1)< \infty$.

\begin{prop}\label{prop_ckp}
Let $A(L)x_{t}=\varepsilon_{t}$ be an \ar\ with $d=1$. Then $\Delta x_t=B(L)\varepsilon_{t}$, where $B(z)=\sum_{n=0}^{\infty}B_{n}z^n$, $z\in \BC$, is such that $\sum_{n=1}^{\infty}n\|B_{n}\|_{\myB}<\infty$ and $\Ima B(1)$ is finite dimensional. The converse does not necessarily hold.
\end{prop}

This shows that $I(1)$ \ars\ necessarily satisfy Assumption 2.1 in \citet{CKP:16}; hence their asymptotic analysis applies and their test can be employed in the present setup.

The next proposition discusses the relation with \citet{HP:WP}, who consider \eqref{eq_AR} with $k=1$ and compact $A^\circ_1$. Similarly, \citet{BSS:17} consider \eqref{eq_AR} with compact $A^\circ_1,\dots,A^\circ_k$ if $k>1$ and \citet{BS:18} consider \eqref{eq_AR} with compact $A^\circ_1,\dots,A^\circ_k$ for $k\geq 1$.

\begin{prop}\label{prop_compact}
Assume that $A^\circ_1,\dots,A^\circ_k$, $k\geq 1$, in \eqref{eq_AR} are compact. Then \eqref{eq_AR} is an \ar.
The converse does not necessarily hold.
\end{prop}

This shows that the present results can be applied to the setups of \citet{HP:WP}, \citet{BSS:17} and \citet{BS:18}. \citet{BSS:17} also consider $x_t=A_1^{\circ}x_{t-1}+\varepsilon_t$ with no compactness assumption on $A_1^{\circ}$, see Proposition \tref{prop_BSS_k_1} below.

Finally, \citet{CHP:WP} consider an error correction form with compact error correction operator and show that in this case the number of $I(1)$ common trends is infinite and the dimension of the \cs\ is finite. This case is not covered by the present results.

\subsection{Example of a non-compact operator}\label{sec_example_intro}

This section illustrates the relevance of Assumption \tref{ass_finite_type} with a simple example. This example is considered again in Section \tref{sec_example} to illustrate the characterization results in the $I(1)$ case.

Consider $x_{t}=A^\circ_1 x_{t-1}+\varepsilon_{t}$ where $A^\circ_1$ is a band operator. Band operators are defined as follows: let $\varphi_1,\varphi_2,\dots$ be an orthonormal basis of $\myH$ and let $(a_{ij})$, where $a_{ij}=\langle A\varphi_j,\varphi_i \rangle$, be the matrix representation of $A\in\myB$ corresponding to $\varphi_1,\varphi_2,\dots$, see e.g. \citet[][Section 2.4]{GGK:03}; $A\in\myB$ is called a band operator if all nonzero entries in its matrix representation $(a_{ij})$ are in a finite number of diagonals parallel to the main diagonal, i.e. there exists an integer $N$ such that $a_{ij}=0$ if $|i-j|>N$, see e.g. \citet[][Section 2.16]{GGK:03}.

Note that a band operator is compact if and only if $\lim_{i,j\rightarrow\infty}a_{ij}=0$, see Theorem 16.4 in \citet[][Section 2.16]{GGK:03}. Here $\lim_{i,j\rightarrow\infty}a_{ij}=0$ is not assumed, hence the operator $A^\circ_1$ is non-necessarily compact. Finally, let $z_{i,t}=\langle \varphi_i,z_t \rangle$ be the $i$-th coordinate of the process $z_t=x_t,\varepsilon_t$, and note that from \eqref{eq_def_Az} one has $A_0 =I- A^\circ_1$, $A_1=A^\circ_{1}$ and $A_n=0$ for $n=2,3,\dots$, in $A(z)=\sum_{n=0}^{\infty}A_{n}(1-z)^n$, so that $A(z)=A_{0}+A_{1}(1-z)$.

Let $(a_{ij})$ be the matrix representation of $A^\circ_1$ and assume that $a_{ij}=0$ for $|i-j|>0$ and $a_{ii}=\alpha_i$, where $\alpha_i\in\BR$, $\alpha_1=1$ and $0<|\alpha_i|<1$, $i=2,3,\dots$, so that $A^\circ_1$ is a band operator. Observe that $x_{t}=A^\circ_1 x_{t-1}+\varepsilon_{t}$ reads
$$
x_{1,t} = x_{1,t-1} + \varepsilon_{1,t},\qquad \qquad x_{i,t} = \alpha_i x_{i,t-1} + \varepsilon_{i,t}, \qquad \qquad i=2,3,\dots.
$$
Remark that $A^\circ_1$ is not compact because $\lim_{i,j\rightarrow\infty}a_{ij}=0$ is not imposed. Next note that $A(z)$ is invertible for all $z\in D(0,\rho)\setminus\{1\}$ for some $\rho>1$ and consider the matrix representation of $A^\circ_{1}=A_1$ and $A_0 =I- A^\circ_1$, i.e.
\begin{equation}\label{eq_I1_mat_repr}
A^\circ_1=A_1=\left(
\begin{array}{ccc}
1 &  &   \\
 & \alpha_2 & \\
& &  \ddots
\end{array}
\right),\qquad A_0=
\left(\begin{array}{ccc}
0 &  &   \\
 & 1-\alpha_2 & \\
& &  \ddots
\end{array}
\right),
\end{equation}
where empty entries are equal to 0, and compute
$$
(\Ima A_0)^\bot=(\csp{\{\varphi_2,\varphi_3,\dots\}})^\bot=\msp{\{\varphi_1\}},\qquad \Ker A_0=\msp{\{\varphi_1\}},
$$
where $\msp\{\cdot\}$ and $\csp\{\cdot\}$ indicate the span of the set of vectors in curly brackets and its closure respectively. Because $0<\dim \Ker A_0=\dim (\Ima A_0)^\bot<\infty$, this shows that $A_0$ is \fof, so that Assumption \tref{ass_finite_type} holds and $x_{t}=A^\circ_1 x_{t-1}+\varepsilon_{t}$ is an \ar\ with non-compact operator.

\section{A characterization of $I(1)$ and $I(2)$ \ars}\label{sec_char_1_2}

This section presents a characterization of $I(1)$ and $I(2)$ \ars. The $I(1)$ case parallels the results in \citet{HP:WP,BSS:17,BS:18}, and it is discussed in Theorem \tref{thm_CHAR_1}. The results for the $I(2)$ case are novel, and they are given in Theorem \tref{thm_CHAR_2}.

%

\subsection{I(1) case}

The following notation is employed: write $A(z)=\sum_{n=0}^{\infty}A_{n}(1-z)^n$ as in \eqref{eq_def_Az} and define
\begin{align}
S_0 &= A_0,&\myalpha_0&=\Ima S_0,&\mybeta_0&=(\Ker S_0)^\bot,\label{eq_def_0}\\
S_1& =P_{\myalpha_0^\bot}A_{1}P_{\mybeta_0^\bot},&\myalpha_1&=\Ima S_1,&\mybeta_1&=(\Ker S_1)^\bot,\label{eq_def_1}
\end{align}
where $P_\eta\in \myB$ indicates the orthogonal projection on $\eta$, i.e. $P_\eta^2=P_\eta$, $\Ima P_\eta=\eta$ and $\Ker P_\eta=\eta^\bot$.

Observe that
$$
\myalpha_1\subseteq \myalpha_0^\bot,\qquad \mybeta_1\subseteq \mybeta_0^\bot
$$
by construction; that is, $\myalpha_1$ is orthogonal to $\myalpha_0$ and $\mybeta_1$ is orthogonal to $\mybeta_0$. Moreover, because 1 is an eigenvalue of finite type, one has $0<\dim \mybeta_0^\bot=\dim \myalpha_0^\bot<\infty$, see Remark \tref{rem_eigFT0}, so that the subspaces $\myalpha_1,\mybeta_1$ are finite dimensional. In the following, $a \rimp b$ indicates that $a$ implies $b$ and the orthogonal direct sum decomposition
\begin{equation}\label{eq_I1_cond}
\myH=\mybeta_0 \oplus \mybeta_1,\qquad \mybeta_1\neq \{ 0 \},
\end{equation}
is called the \pole{1} condition.

\begin{thm}[A characterization of $I(1)$ \ars]\label{thm_CHAR_1}
Consider an \ar\ $A(L)x_{t}=\varepsilon_{t}$ and let $\mybeta_0$, $\mybeta_1$ be as in \eqref{eq_def_0}, \eqref{eq_def_1} respectively. Then $x_{t}$ is $I(1)$ if and only if the \pole{1} condition in \eqref{eq_I1_cond} holds; in this case, the common trends representation of $x_t$ is found by setting $d=1$ in \eqref{eq_CT}. Moreover, $
\Ima C_0=\mybeta_1$ is the finite dimensional \as, $\mybeta_0$ is the infinite dimensional \cs\ and for any nonzero $v\in \myH$ one has
\begin{align}
&&  v\in \mybeta_0 && \rimp && \langle v,x_t\rangle \sim I(0),&& \label{eq_char_I1_0}\\
&& v\in \mybeta_1 && \rimp && \langle v,x_t\rangle \sim I(1),&& \label{eq_char_I1+1}
\end{align}
where $\mybeta_1=\mybeta_0^\bot\neq \{ 0 \}$.
\end{thm}
Theorem \tref{thm_CHAR_1} shows that an \ar\ generates an
$I(1)$ process if and only if $\mybeta_1 = \mybeta_0^\bot\neq \{ 0 \}$. The common trends representation of $x_t$ shows that the $I(1)$ stochastic trends $s_{1,t}$ are loaded into the process by $C_0$; because $\Ima C_0$ coincides with $\mybeta_1$, $\mybeta_1$ is the finite dimensional \as\ and the number of $I(1)$ trends in $x_t$ is finite and equal to $\dim \mybeta_1$.

Moreover, because $\Ima C_0=\mybeta_1 = \mybeta_0^\bot$, for any nonzero $v \in \mybeta_0$ one has $\langle v,C_0 y \rangle = 0$ for all $y\in \myH$, and hence also for $y=s_{d,t}$ in \eqref{eq_CT}; this implies that $\langle v,x_t\rangle$ is stationary, i.e. $\mybeta_0$ is the infinite dimensional \cs.
Note that this decomposition is orthogonal.

Using orthogonal projections, one can see that this orthogonal direct sum decomposition can be employed in general to characterize the degree of integration of any $v$-characteristic of the process.
In fact, note that \eqref{eq_I1_cond} implies $P_{\tau_0}+P_{\tau_1}=I$, where $P_{\tau_h}$ is the orthogonal projection onto $\tau_h$; hence for any nonzero $v\in \myH$ one has $\langle v,x_t\rangle=\langle v_0,x_t\rangle+\langle v_1,x_t\rangle$, where $v_h=P_{\tau_h}v\in \tau_h$, so that \eqref{eq_char_I1_0} and \eqref{eq_char_I1+1} describe the order of integration of any nonzero $v$-characteristic $\langle v,x_t\rangle$ of $x_t$. In particular one has $\langle v,x_t\rangle\sim I(1)$ if and only if $v_1\neq 0$, because $\langle v_0,x_t\rangle$ is $I(0)$.

Theorem \ref{thm_CHAR_1} further shows that for any nonzero $ v\in \mybeta_0$, $\langle v,x_t\rangle$ is not only stationary, but $I(0)$. This echoes the finite dimensional case, see Theorem 4.2 in \citet{Joh:96}, except for the fact that the number of $I(0)$ cointegrating relations is infinite.

\begin{remark}\label{rem_I1_Joh}
Let $\msp\{a\}$ indicate $\msp\{a_1,\dots, a_k\}$ when its argument $a$ is a matrix with $k$ columns $a_i$, $a=(a_1,\dots,a_k)$.
In the finite dimensional case $\myH=\BR^p$, \citet{FP:16} show that the $I(1)$ condition in Theorem 4.2 in \citet{Joh:96} can be equivalently stated as $\BR^p=\myalpha_0 \oplus \myalpha_1=\mybeta_0 \oplus \mybeta_1$, $\myalpha_1\neq \{ 0 \}$ and $\mybeta_1\neq \{ 0 \}$, where $\myalpha_{h}=\msp\{\alpha_h\}$, $\mybeta_{h}=\msp\{\beta_h\}$, $h=0,1$, and
the bases $\alpha_h$, $\beta_h$ are defined by the \rf s $A_{0}=\alpha_{0}\beta_{0}'$ and
$P_{\myalpha_0^\bot}A_{1}P_{\mybeta_0^\bot}=\alpha _{1}\beta_{1}'$, i.e. $\alpha_h$, $\beta_h$ are full-column-rank matrices that respectively span the column space $\myalpha_h$ and the row space $\mybeta_h$ of the corresponding matrix. Except for the fact that $\dim \myalpha_0=\dim \mybeta_0$ is finite when $\myH=\BR^p$, this mirrors what happens in the present infinite dimensional case.
\end{remark}

\begin{remark}\label{rem_I1_iota}
The \pole{1} condition in \eqref{eq_I1_cond} is equivalent to $\mybeta_1=\mybeta_0^\bot\neq \{ 0 \}$. Moreover, Theorem \tref{thm_pole_order} in Appendix \tref{app_lemmas} shows that it can be equivalently stated as $(i)$ $\myH=\myalpha_0 \oplus \myalpha_1$, $\myalpha_1\neq \{ 0 \}$, $(ii)$ $\myalpha_1=\myalpha_0^\bot\neq \{ 0 \}$, $(iii)$ $\Ima C_0=\mybeta_1$, $(iv)$ $\Ker C_0=\myalpha_0$.
\end{remark}

The \pole{1} condition is next compared to equivalent conditions in the literature. \citet[][Definition 4.3]{BSS:17} define the following non-orthogonal direct sum decomposition
\begin{equation}\label{eq_BSS_I1}
\myH=\Ima A_0 \oplus A_1 \Ker A_0,
\end{equation}
where $A_0,A_1$ are as in \eqref{eq_def_Az}; they call \eqref{eq_BSS_I1} the `Johansen $I(1)$ condition'. Their Theorem 4.1 assumes that $x_t=A_1^{\circ}x_{t-1}+\varepsilon_t$ with no compactness assumption on $A_1^{\circ}$ and that the $k=1$ version of \eqref{eq_BSS_I1} holds, i.e. $\myH=\Ima A_0 \oplus \Ker A_0$.
Under these conditions, they finds the common trends representation \eqref{eq_CT} with $d=1$ and $\Ima C_0 = \Ker A_0$.
The following proposition clarifies the connection between \ars\ and their result.

\begin{prop}\label{prop_BSS_k_1}
Consider $x_t=A_1^{\circ}x_{t-1}+\varepsilon_t$ with no compactness assumption on $A_1^{\circ}$ and let $\myH=\Ima A_0 \oplus \Ker A_0$. If $\Ker A_0$ is finite dimensional then $x_t=A_1^{\circ}x_{t-1}+\varepsilon_t$ is an \ar.
\end{prop}
One can observe that
the case with infinite dimensional $\Ker A_0$, which correspons to an infinite dimensional \as, is not covered by the present results.

Finally, the following proposition proves the equivalence of the orthogonal direct sum condition in \eqref{eq_I1_cond} and the nonorthogonal direct sum conditions in \eqref{eq_BSS_I1}.

\begin{prop}\label{prop_equiv_BSS}
Let $A(L)x_{t}=\varepsilon_{t}$ be an \ar; then the $I(1)$ condition in \eqref{eq_BSS_I1} is equivalent to the \pole{1} condition in \eqref{eq_I1_cond}.
\end{prop}

\subsection{I(2) case}

The $I(2)$ case is considered next. Consider $A(z)=\sum_{n=0}^{\infty}A_{n}(1-z)^n$ in \eqref{eq_def_Az}, and let $\myalpha_0,\mybeta_0$ be as in \eqref{eq_def_0}, consider $\myalpha_1,\mybeta_1$ as in \eqref{eq_def_1}, and define
\begin{equation}\label{eq_def_2}
S_2=P_{\mya_{2}^\bot}A_{2,1}P_{\myb_{2}^\bot},\qquad \myalpha_2=\Ima S_2,\qquad \mybeta_2=(\Ker S_2)^\bot,
\end{equation}
where $\mya_{2}=\myalpha_0\oplus\myalpha_1$, $\myb_{2}=\mybeta_0\oplus\mybeta_1$ and $A_{2,1}=A_2-A_1A_0^+A_1$, where the generalized inverse $A_0^+$ exists and it is unique, see Remark \tref{rem_eigFT0}.

Observe that
$$
\myalpha_2\subseteq (\myalpha_0\oplus\myalpha_1)^\bot,\qquad \mybeta_2\subseteq (\mybeta_0\oplus\mybeta_1)^\bot
$$
by construction; that is, for $0< j<h$, $\myalpha_h$ is orthogonal to $\myalpha_j$, and $\mybeta_h$ is orthogonal to $\mybeta_j$. Moreover, because $0<\dim \myalpha_0^\bot = \dim \mybeta_0^\bot < \infty$, the subspaces $\myalpha_2,\mybeta_2$ are finite dimensional. In the following, the orthogonal direct sum decomposition
\begin{equation}\label{eq_I2_cond}
\myH=\mybeta_0 \oplus \mybeta_1\oplus \mybeta_2,\qquad \mybeta_2\neq \{ 0 \},
\end{equation}
is called the \pole{2} condition.

\begin{thm}[A characterization of $I(2)$ \ars]\label{thm_CHAR_2}
Consider an \ar\ $A(L)x_{t}=\varepsilon_{t}$, let $\mybeta_0$, $\mybeta_1$, $\mybeta_2$ be as in \eqref{eq_def_0}, \eqref{eq_def_1}, \eqref{eq_def_2} respectively and let $A_0^+$ be the generalized inverse of $A_0$; then $x_{t}$ is $I(2)$ if and only if the \pole{2} condition in \eqref{eq_I2_cond} holds. In this case, the common trends representation of $x_t$ is found by setting $d=2$ in \eqref{eq_CT}. Moreover, $
\Ima C_0=\mybeta_2$ is the finite dimensional \as, $\mybeta_0\oplus\mybeta_1$ is the infinite dimensional \cs\ and for any
nonzero $v$-characteristics $v\in\myH$ one has
\begin{align}
&&v\in \mybeta_0 && \rimp &&  \langle v,x_t\rangle + \langle v,A_0^+A_{1}\Delta x_t\rangle \sim I(0),&& \label{eq_char_I2+0}\\
&&
v\in \mybeta_1 && \rimp &&  \langle v,x_t\rangle \sim I(1),&& \label{eq_char_I2+1}\\
&&
v\in \mybeta_2 && \rimp && \langle v,x_t\rangle \sim I(2),&& \label{eq_char_I2+2}
\end{align}
where $\mybeta_1\subset\mybeta_0^\bot$ and $\mybeta_2=(\mybeta_0\oplus\mybeta_1)^\bot\neq \{ 0 \}$.
\end{thm}

Some remarks on Theorem \tref{thm_CHAR_2} are in order.

\begin{remark}\label{rem-one}An \ar\ generates an $I(2)$ process if and only if $\mybeta_2 = (\mybeta_0\oplus\mybeta_1)^\bot\neq \{ 0 \}$. The common trends representation of $x_t$ shows that the $I(2)$ stochastic trends $s_{2,t}$ are loaded into the process by $C_0$; because $\Ima C_0$ coincides with $\mybeta_2$, $\mybeta_2$ is the finite dimensional \as\ and the number of $I(2)$ trends in $x_t$ is finite and equal to $\dim \mybeta_2$.
\end{remark}

\begin{remark}\label{rem-two}
Moreover, because $\Ima C_0=\mybeta_2=(\mybeta_0 \oplus \mybeta_1)^\bot$, for any nonzero $v \in \mybeta_0 \oplus \mybeta_1$
one has $\langle v,C_0 y \rangle = 0$ for all $y\in \myH$, and hence also for $y=s_{d,t}$ in \eqref{eq_CT}; this implies that $\langle v,x_t\rangle$ is at most $I(1)$, i.e. $\mybeta_0 \oplus \mybeta_1$ is the infinite dimensional \cs.
Note that this decomposition is orthogonal. Using orthogonal projections, one can see that this orthogonal direct sum decomposition can be employed in general to characterize the degree of integration of any $v$-characteristic of the process.
In fact, note that \eqref{eq_I2_cond} implies $P_{\tau_0}+P_{\tau_1}+P_{\tau_2}=I$, where $P_{\tau_h}$ is the orthogonal projection onto $\tau_h$; hence for any nonzero $v\in \myH$ one has $\langle v,x_t\rangle=\langle v_0,x_t\rangle+\langle v_1,x_t\rangle+\langle v_2,x_t\rangle$, where $v_h=P_{\tau_h}v\in \tau_h$, so that \eqref{eq_char_I2+0}, \eqref{eq_char_I2+1} and \eqref{eq_char_I2+2} describe the order of integration of any nonzero $v$-characteristic $\langle v,x_t\rangle$ of $x_t$.
In particular, one has $\langle v,x_t\rangle\sim I(2)$ if and only if $v_2\neq  0 $, because $\langle v_0,x_t\rangle+\langle v_1,x_t\rangle$ is at most $I(1)$.
\end{remark}
\begin{remark}\label{rem-three}
Theorem \ref{thm_CHAR_2} further shows that in $\mybeta_0$, which is infinite dimensional, one finds the cointegrating vectors that allow for polynomial cointegration of order 0 and in $\mybeta_1$, with
$0 \leq \dim \mybeta_1 < \infty$, those that don't allow for polynomial cointegration. Specifically, any nonzero $v_0\in \mybeta_0$, if one combines levels and first differences as in $\langle v_0,x_t\rangle + \langle v_0,A_0^+A_{1}\Delta x_t\rangle$ one finds an $I(0)$ process; given that $\langle v_0,A_0^+A_{1}\Delta x_t\rangle$ can as well be equal to 0, there may exist a nonzero $v_0\in \mybeta_0$ such that $\langle v_0,x_t\rangle\sim I(0)$. This cannot happen in the $\mybeta_1$ subspace, in which every nonzero $v_1\in \mybeta_1$ is such that $\langle v_1,x_t\rangle\sim I(1)$. Apart from the fact that the number of $I(0)$ cointegrating relations is infinite, this mimics the finite dimensional case, see Theorem 4.6 in \citet{Joh:96}.
\end{remark}
\begin{remark}\label{rem_I2_Joh}
In the finite dimensional case $\myH=\BR^p$, \citet{FP:16} show that the $I(2)$ condition in Theorem 4.6 in \citet{Joh:96} can be equivalently stated as $\BR^p=\myalpha_0 \oplus \myalpha_1\oplus \myalpha_2=\mybeta_0 \oplus \mybeta_1\oplus \mybeta_2$, $\myalpha_2\neq \{ 0 \}$ and $\mybeta_2\neq \{ 0 \}$, where $\myalpha_{h}=\msp\{\alpha_h\}$, $\mybeta_{h}=\msp\{\beta_h\}$, $h=0,1,2$, and the bases $\alpha_h$, $\beta_h$ are defined by the \rf s $A_{0}=\alpha_{0}\beta_{0}'$, $P_{\myalpha_0^\bot}A_{1}P_{\mybeta_0^\bot}=\alpha _{1}\beta_{1}'$ and $P_{\mya_{2}^\bot}A_{2,1}P_{\myb_{2}^\bot}=\alpha_{2}\beta_{2}'$ where 
$A_{2,1}=A_2-A_1\bar{\beta}_0\bar{\alpha}_0'A_1$,
$(\alpha_0 \beta_0 ' )^+ = \bar{\beta}_0 \bar{\alpha}_0'$ and $\bar{\eta}=\eta(\eta'\eta)^{-1}$ for a generic full-column-rank matrix $\eta$. Again here, apart from the fact that $\dim \myalpha_0=\dim \mybeta_0$ is finite when $\myH=\BR^p$, this is exactly what happens in the infinite dimensional case.
\end{remark}

\begin{remark}\label{rem_I2_iota}
The \pole{2} condition in \eqref{eq_I2_cond} is equivalent to $\mybeta_2=(\mybeta_0\oplus\mybeta_1)^\bot\neq \{ 0 \}$. Moreover, Theorem \tref{thm_pole_order} in Appendix \tref{app_lemmas} shows that it can be equivalently stated as $(i)$ $\myH=\myalpha_0 \oplus \myalpha_1\oplus \myalpha_2$, $\myalpha_2\neq \{ 0 \}$, $(ii)$ $\myalpha_2=(\myalpha_0\oplus\myalpha_1)^\bot\neq \{ 0 \}$, $(iii)$ $\Ima C_0=\mybeta_2$, $(iv)$ $\Ker C_0=\myalpha_0\oplus\myalpha_1$.
\end{remark}

\subsection{Illustrations}\label{sec_example}

This section illustrates Theorems \tref{thm_CHAR_1} and \tref{thm_CHAR_2} via two simple examples, called the $I(1)$ and the
$I(2)$ examples.

\textbf{$I(1)$ example.}
Consider the setup in Section \tref{sec_example_intro}. Here the analysis should deliver that $x_t$ is $I(1)$, the \as\ coincides with $\msp\{\varphi_1\}$ and the \cs\ with $\csp\{\varphi_2,\varphi_3,\dots\}$. Since $\langle v,x_t\rangle$ is $I(0)$ for any nonzero $v\in \csp\{\varphi_2,\varphi_3,\dots\}$ and $\langle v,x_t\rangle$ is $I(1)$ for any nonzero $v\in \msp\{\varphi_1\}$, the analysis should further convey
that $\mybeta_0=\csp\{\varphi_2,\varphi_3,\dots\}$ and $\mybeta_1=\msp\{\varphi_1\}$.

From \eqref{eq_I1_mat_repr}, one has
\begin{align*}
\myalpha_0=\Ima A_0=\csp{\{\varphi_2,\varphi_3,\dots\}},& & & & & &\mybeta_0=(\Ker A_0)^\bot=(\msp{\{\varphi_1\}})^\bot=\csp{\{\varphi_2,\varphi_3,\dots\}},\\
\myalpha_1=\Ima P_{\myalpha_0^\bot}A_{1}P_{\mybeta_0^\bot} = \msp{\{\varphi_1\}},& & & & & & \mybeta_1=(\Ker P_{\myalpha_0^\bot}A_{1}P_{\mybeta_0^\bot})^\bot=(\csp{\{\varphi_2,\varphi_3,\dots\}})^\bot=\msp{\{\varphi_1\}}.
\end{align*}

This shows that $\myH=\mybeta_0 \oplus \mybeta_1$, $\mybeta_1\neq \{ 0 \}$, so that the \pole{1} condition in \eqref{eq_I1_cond} holds and Theorem \tref{thm_CHAR_1} applies: the common trends representation of $x_t$ is found by setting $d=1$ in \eqref{eq_CT}, $\Ima C_0=\mybeta_1=\msp{\{\varphi_1\}}$ is the finite dimensional \as\ and $\mybeta_0=\csp{\{\varphi_2,\varphi_3,\dots\}}$ is the infinite dimensional \cs.

\textbf{$I(2)$ example.} 
Let $(a_{ij})$ be the matrix representation of $A^\circ_1$ and assume that $a_{ij}=0$ for $|i-j|>1$, $a_{12}=1$ and $a_{ii}=\alpha_i$, where $\alpha_i\in\BR$, $\alpha_1=\alpha_2=\alpha_3=1$ and $0<|\alpha_i|<1$, $i=4,5,\dots$. Again here, $A^\circ_1$ is not necessarily compact but $x_{t}=A^\circ_1 x_{t-1}+\varepsilon_{t}$ is an \ar, as shown below. Observe that $x_{t}=A^\circ_1 x_{t-1}+\varepsilon_{t}$ reads
$$
\begin{array}{lll}
x_{1,t} = x_{1,t-1} + x_{2,t-1} + \varepsilon_{1,t},& \qquad \qquad x_{2,t} = x_{2,t-1} + \varepsilon_{2,t},& \qquad \qquad x_{3,t} =  x_{3,t-1}+ \varepsilon_{3,t}, \\
x_{i,t} = \alpha_i x_{i,t-1} + \varepsilon_{i,t}, & \qquad \qquad i=4,5,\dots.
\end{array}
$$
Hence the analysis should deliver that $x_t$ is $I(2)$, the \as\ coincides with $\msp\{\varphi_1\}$ and the \cs\ with $\csp\{\varphi_2,\varphi_3,\dots\}$. Next note that $\langle v,x_t\rangle$ is $I(0)$ for any nonzero $v\in \csp\{\varphi_4,\varphi_5,\dots\}$ and $\langle v,x_t\rangle$ is $I(1)$ for any nonzero $v\in \msp\{\varphi_2,\varphi_3\}$. Moreover, because $\Delta x_{1,t} = x_{2,t-1} + \varepsilon_{1,t} = x_{2,t} - \varepsilon_{2,t} + \varepsilon_{1,t}$, one has that $x_{2,t} - \Delta x_{1,t}$ is $I(0)$, i.e. $\langle \varphi_2 , x_{t}\rangle - \langle \varphi_1 , \Delta x_{t}\rangle$ is $I(0)$, so that $\langle \varphi_2 , x_{t}\rangle$ allows for polynomial cointegration while $\langle \varphi_3 , x_{t}\rangle$ does not. Hence the analysis should further convey
that $\mybeta_0=\csp\{\varphi_2,\varphi_4,\varphi_5,\dots\}$, $\mybeta_1=\msp\{\varphi_3\}$, $\mybeta_2=\msp\{\varphi_1\}$, $\langle \varphi_2,A_0^+A_1\Delta x_{i,t}\rangle=-\Delta x_{i,t}$, and $\langle \varphi_i,A_0^+A_1\Delta x_{i,t}\rangle=0$ for $i=4,5,\dots$.

Consider the matrix representation of $A^\circ_{1}=A_1$ and $A_0 =I- A^\circ_1$,
i.e.
$$
A^\circ_1=A_1=\left(
\begin{array}{ccccc}
1 & 1 &  &  &  \\
  & 1 &  &  & \\
  &   &1 &  &  \\
  &   &  & \alpha_4 &   \\
  &   &  &          & \ddots
\end{array}
\right),\qquad A_0=\left(
\begin{array}{ccccc}
0 & -1 &  &  &  \\
  & 0 &  &  & \\
  &   &0 &  &  \\
  &   &  & 1-\alpha_4 &   \\
  &   &  &          & \ddots
\end{array}
\right),
$$
where empty entries are equal to 0. Compute
$$
\myalpha_0=\Ima A_0=\csp{\{\varphi_1,\varphi_4,\varphi_5,\dots\}},\qquad \mybeta_0=(\Ker A_0)^\bot=(\msp{\{\varphi_1,\varphi_3\}})^\bot=\csp{\{\varphi_2,\varphi_4,\varphi_5,\dots\}},
$$
so that $\myalpha_0^\bot=\msp{\{\varphi_2,\varphi_3\}}$ and $\mybeta_0^\bot=\msp{\{\varphi_1,\varphi_3\}}$; because $0<\dim \Ker A_0=\dim (\Ima A_0)^\bot<\infty$, this shows that $A_0$ is \fof\ and because $A(z)=I-A^\circ_1 z$ is invertible for all $z\in D(0,\rho)\setminus\{1\}$ for some $\rho>1$, $x_{t}=A^\circ_1 x_{t-1}+\varepsilon_{t}$ is an \ar\ with non-compact operator.

Next compute
$$
\myalpha_1=\Ima P_{\myalpha_0^\bot}A_{1}P_{\mybeta_0^\bot} = \msp{\{\varphi_3\}},\qquad \mybeta_1=(\Ker P_{\myalpha_0^\bot}A_{1}P_{\mybeta_0^\bot})^\bot=(\csp{\{\varphi_1,\varphi_2,\varphi_4,\varphi_5,\dots\}})^\bot=\msp{\{\varphi_3\}}.
$$
This shows that $\mybeta_1\subset \mybeta_0^\bot$, so that the \pole{1} condition in \eqref{eq_I1_cond} does not hold and the process is $I(d)$ for some finite $d=2,3,\dots$.

Now consider $P_{\mya_{2}^\bot}A_{2,1}P_{\myb_{2}^\bot}$ in \eqref{eq_def_2}; since $\mya_{2}=\myalpha_0\oplus\myalpha_1=\csp{\{\varphi_1,\varphi_3,\varphi_4,\dots\}}$ and $\myb_{2}=\mybeta_0\oplus\mybeta_1=\csp{\{\varphi_2,\varphi_3,\dots\}}$, one has $\mya_{2}^\bot=\msp{\{\varphi_2\}}$ and $\myb_{2}^\bot=\msp{\{\varphi_1\}}$. Note that $A_2=0$ and hence $A_{2,1}=-A_1A_0^+A_1$; thus $P_{\mya_{2}^\bot}A_{2,1}P_{\myb_{2}^\bot}=-P_{\msp\{\varphi_2\}}A_1A_0^+A_1
P_{\msp\{\varphi_1\}}$
and because $P_{\msp\{\varphi_2\}}A_1=P_{\msp\{\varphi_2\}}$ and $A_1P_{\msp\{\varphi_1\}}=P_{\msp\{\varphi_1\}}$, one has $P_{\mya_{2}^\bot}A_{2,1}P_{\myb_{2}^\bot}=-P_{\msp\{\varphi_2\}}A_0^+P_{\msp \{\varphi_1\}}$. Next the matrix representation of $A_0^+$ is investigated; from Lemma \tref{lem_gen_inv_h} one has $\Ker A_0^+ =(\Ima A_0)^\bot$, $A_0^+ A_0=P_{(\Ker A_0)^\bot}$ and because $(\Ima A_0)^\bot=\msp{\{\varphi_2,\varphi_3\}}$ and $(\Ker A_0)^\bot=\csp{\{\varphi_2,\varphi_4,\varphi_5,\dots\}}$ one has
$$
A_0^+=\left(
\begin{array}{ccccc}
0 &  &  &  &  \\
-1  & 0 &  &  & \\
  &   &0 &  &  \\
  &   &  & \frac{1}{1-\alpha_4} &   \\
  &   &  &          & \ddots
\end{array}
\right).
$$
This implies that the only nonzero element in the matrix representation of $P_{\mya_{2}^\bot}A_{2,1}P_{\myb_{2}^\bot}=-P_{\msp\{\varphi_2\}}A_0^+P_{\msp\{\varphi_1\}}$ is a one in row 2 and column 1, so that
$$
\myalpha_2=\Ima P_{\mya_{2}^\bot}A_{2,1}P_{\myb_{2}^\bot}=\msp{\{\varphi_2\}},\qquad \mybeta_2=(\Ker P_{\mya_{2}^\bot}A_{2,1}P_{\myb_{2}^\bot})^\bot=(\csp{\{\varphi_2,\varphi_3,\dots\}})^\bot=
\msp{\{\varphi_1\}}.
$$
Hence $\myH=\mybeta_0 \oplus \mybeta_1 \oplus \mybeta_2$, $\mybeta_2\neq \{ 0 \}$, i.e. the \pole{2} condition in \eqref{eq_I2_cond} holds and Theorem \tref{thm_CHAR_2} applies: the common trends representation of $x_t$ is found by setting $d=2$ in \eqref{eq_CT}, $\Ima C_0=\mybeta_2=\msp{\{\varphi_1\}}$ is the finite dimensional \as\ and $\mybeta_0 \oplus \mybeta_1=\csp{\{\varphi_2,\varphi_3,\dots\}}$ is the infinite dimensional \cs. Moreover, for any nonzero $v\in\myH$ one has
\begin{align*}
&&v\in \mybeta_0=\csp{\{\varphi_2,\varphi_4,\varphi_5,\dots\}} && \rimp &&  \langle v,x_t\rangle + \langle v,A_0^+A_{1}\Delta x_t\rangle \sim I(0),&&\\
&&v\in \mybeta_1=\msp{\{\varphi_3\}} && \rimp &&  \langle v,x_t\rangle \sim I(1),&&\\
&&v\in \mybeta_2=\msp{\{\varphi_1\}} && \rimp && \langle v,x_t\rangle \sim I(2).&&
\end{align*}
Note that $\mybeta_1\subset\mybeta_0^\bot$ and $\mybeta_2=(\mybeta_0\oplus\mybeta_1)^\bot\neq \{ 0 \}$. Moreover, $\langle \varphi_2,A_0^+A_{1}\Delta x_t\rangle=-\Delta x_{1,t}-\Delta x_{2,t}$ and $\langle \varphi_i,A_0^+A_{1}\Delta x_{t}\rangle=\frac{\alpha_i}{1-\alpha_i} \Delta x_{i,t}$, for $i=4,5,\dots$; hence $\langle \varphi_2,x_t\rangle + \langle \varphi_2,A_0^+A_{1}\Delta x_t\rangle=
x_{2,t}-\Delta x_{1,t}-\Delta x_{2,t}$
and, for $i=4,5,\dots$, $\langle \varphi_i,x_t\rangle + \langle \varphi_i,A_0^+A_{1}\Delta x_t\rangle=
x_{i,t} + \frac{\alpha_i}{1-\alpha_i} \Delta x_{i,t}$. This shows that $\langle v,x_t\rangle + \langle v,A_0^+A_{1}\Delta x_t\rangle$ contains stationary terms ($\Delta x_{2,t}$ and $\frac{\alpha_i}{1-\alpha_i} \Delta x_{i,t}$) that are not necessary for cointegration; these can be eliminated by considering $\langle v,A_0^+A_{1}P_{\mybeta_2}\Delta x_t\rangle$ instead of $\langle v,A_0^+A_{1}\Delta x_t\rangle$, as in the finite dimensional $I(2)$ case, see Theorem 4.6 in \citet{Joh:96}.

\section{A characterization of $I(d)$ \ars}\label{sec_char_d}

This section extends the results in Section \tref{sec_char_1_2} to the general $I(d)$, $d=1,2,\dots<\infty$, case. Theorem \tref{thm_CHAR_d} provides a necessary and sufficient condition for \ars\ to be $I(d)$ and it is shown that under this condition the space $\myH$ is decomposed into the direct sum of $d+1$ orthogonal subspaces $\mybeta_h$, $\myH=\mybeta_0\oplus\mybeta_1\oplus\dots\oplus\mybeta_d$, $\mybeta_d\neq \{ 0 \}$, that are defined in terms of $A_0,A_1,\dots, A_d$ in \eqref{eq_def_Az}, see Definition \tref{def_LRF} below.

The finite dimensional \as\ coincides with $\mybeta_d$ and $\mybeta_0\oplus \mybeta_1\oplus\dots\oplus\mybeta_{d-1}$ is the infinite dimensional \cs. In $\mybeta_0$, which is infinite dimensional, one finds the cointegrating vectors that allow for polynomial cointegration of order 0 and in $\mybeta_h$, $h=1,\dots,d-2$, which is finite dimensional and can as well be equal to 0, those that allow for polynomial cointegration of order $h$. In $\mybeta_{d-1}$, with $ 0 \leq \dim \mybeta_{d-1} < \infty$, those that are $I(d-1)$ and don't allow for polynomial cointegration. Finally, any nonzero $v\in\mybeta_{d}$ is such that $\langle v,x_t \rangle \sim I(d)$. The results in Section \tref{sec_char_1_2} are found as special cases for $d=1$ and $d=2$. Before stating the results, some definitions are introduced.
\begin{defn}[$S_h,\myalpha_h,\mybeta_h$, and $A_{h,n}$]\label{def_LRF}
Consider an \ar\ $A(L)x_{t}=\varepsilon_{t}$, where $A(z)=\sum_{n=0}^{\infty}A_{n}(1-z)^n$ is as in \eqref{eq_def_Az}. Let
$$
S_0=A_0,\qquad \myalpha_0=\Ima S_0,\qquad \mybeta_0=(\Ker S_0)^\bot
$$
and for $h=1,2,\dots$ define
\begin{equation}\label{eq_Sh}
S_h=P_{\mya_{h}^\bot}A_{h,1}P_{\myb_{h}^\bot},\qquad \myalpha_h=\Ima S_h,\qquad \mybeta_h=(\Ker S_h)^\bot,
\end{equation}
where
\begin{equation}\label{eq_Ih_Kh}
\mya_{h}=\myalpha_0\oplus\dots\oplus\myalpha_{h-1},\qquad \myb_{h}=\mybeta_0\oplus\dots\oplus\mybeta_{h-1}
\end{equation}
and
\begin{equation}\label{eq_Ahn}
A_{h,n} = \left\{
\begin{array}{cl}
A_{n} & \mbox{ for } h=1 \\
A_{h-1,n+1}-A_{h-1,1}\sum_{j=0}^{h-2} S_j^+ A_{j+1,n} & \mbox{ for } h=2,3,\dots
\end{array}
\right.,\qquad n=1,2,\dots .
\end{equation}
\end{defn}

\smallskip A few remarks are in order.

\begin{remark}\label{rem_LRF}
First note that for $h=1,2$ \eqref{eq_Sh}, \eqref{eq_Ih_Kh} and \eqref{eq_Ahn} deliver \eqref{eq_def_1} and \eqref{eq_def_2} respectively. Next observe that for $h=1,2,\dots$ one has
\begin{equation}\label{eq_subset_mybeta}
\myalpha_h\subseteq (\myalpha_0\oplus \cdots \oplus \myalpha_{h-1})^\bot,\qquad \mybeta_h\subseteq (\mybeta_0\oplus \cdots \oplus \mybeta_{h-1})^\bot
\end{equation}
by construction; that is, for $0<j<h$, $\myalpha_h$ is orthogonal to $\myalpha_j$ and $\mybeta_h$ is orthogonal to $\mybeta_j$. Moreover, because $0<\dim \myalpha_0^\bot = \dim \mybeta_0^\bot < \infty$, for $h=1,2,\dots$ the subspaces $\myalpha_h$ and $\mybeta_h$ are finite dimensional and possibly of dimension equal to 0.

Note also that, as $h$ increases, the finite dimensional subspaces $\mya_{h}^\bot=(\myalpha_0\oplus\dots\oplus\myalpha_{h-1})^\bot$ and $\myb_{h}^\bot=(\mybeta_0\oplus\dots\oplus\mybeta_{h-1})^\bot$ have non-increasing dimension and, because $0<\dim \myalpha_0^\bot = \dim \mybeta_0^\bot < \infty$, they will eventually have dimension $0$. This shows that only a finite number of $\myalpha_h,\mybeta_h$ are nonzero. Let $s$ be the value of $h$ such that $\mya_{s}^\bot\neq \{ 0 \}$, $\myb_{s}^\bot\neq \{ 0 \}$ and $\mya_{s}^\bot=\myb_{s}^\bot=\{ 0 \}$. 
As shown in Theorem \tref{thm_pole_order} in Appendix \tref{app_lemmas}, the integer $s$ is precisely the order of the pole of $A(z)^{-1}$ at $z=1$.

Finally observe that the generalized inverse of $S_h$, $S^+_h$, exists and it is unique for $h=0,1,\dots$, because $\Ima S_h$, $h=0,1,\dots$, is closed; in fact, $S_0$ is \fof, see Remark \tref{rem_eigFT0}, and $\dim \Ima S_h<\infty$ for $h=1,2,\dots$.
\end{remark}

In the following, the orthogonal direct sum decomposition
\begin{equation}\label{eq_Id_cond}
\myH=\mybeta_0 \oplus \mybeta_1 \oplus \cdots \oplus \mybeta_d,\qquad \mybeta_d\neq \{ 0 \},
\end{equation}
is called the \pole{d} condition.

\begin{thm}[A characterization of $I(d)$ \ars]\label{thm_CHAR_d}
Consider an \ar\ $A(L)x_{t}=\varepsilon_{t}$
and let $S_h$, $\mybeta_h$ and $A_{h,n}$ be as in Definition \tref{def_LRF}. Then $x_{t}$ is $I(d)$ if and only if the \pole{d} condition in \eqref{eq_Id_cond} holds. In this case, the common trends representation of $x_t$ is found in \eqref{eq_CT}. Moreover, $\Ima C_0=\mybeta_d$ is the finite dimensional \as, $\mybeta_0 \oplus \mybeta_1 \oplus \cdots \oplus \mybeta_{d-1}$ is the infinite dimensional \cs\ and for any nonzero $v$-characteristic $v\in \myH$ and for $h=0,1,\dots,d$, one has
\begin{equation}\label{eq_char_Id}
v\in \mybeta_{h}\qquad \rimp \qquad \langle v,x_t\rangle + \sum_{n=1}^{d-h-1}\langle v,S_h^+A_{h+1,n}\Delta^n x_t\rangle \sim I(h),
\end{equation}
where empty sums are defined to be $0$, $\mybeta_h\subset (\mybeta_0\oplus \cdots \oplus \mybeta_{h-1})^\bot$ for $h=1,\dots,d-1$ and $\mybeta_d = (\mybeta_0\oplus \cdots \oplus \mybeta_{d-1})^\bot\neq \{ 0 \}$.
\end{thm}
\begin{remark}\label{rem-11}
Theorem \tref{thm_CHAR_d} provides a full description of the properties of an $I(d)$ \ar\ for a generic $d=1,2,\dots<\infty$. For $d=1$ and $d=2$ one finds the most empirically relevant $I(1)$ and $I(2)$ cases discussed in Theorems \tref{thm_CHAR_1}, \tref{thm_CHAR_2}. Remark that all the relevant quantities in Theorem \tref{thm_CHAR_d} are expressed in terms of the AR operators via Definition \tref{def_LRF}.
\end{remark}
\begin{remark}\label{rem-12}
Also note that \eqref{eq_char_Id} provides information that parallels the Triangular Representation for finite dimensional process discussed in \citet{Phi:91} and \citet{SW:93}; see also \citet[][Corollary 4.6]{FP_idcoeff}.
\end{remark}
\begin{remark}\label{rem-13}
An \ar\ generates an $I(d)$ process if and only if $\mybeta_d = (\mybeta_0\oplus\cdots\oplus\mybeta_{d-1})^\bot \neq \{ 0 \}$. The common trends representation of $x_t$ shows that the $I(d)$ stochastic trends $s_{d,t}$ are loaded into the process by $C_0$; because $\Ima C_0$ coincides with $\mybeta_d$, $\mybeta_d$ is the finite dimensional \as\ and the number of $I(d)$ trends in $x_t$ is finite and equal to $\dim \mybeta_d$.

Moreover, because $\Ima C_0=\mybeta_d = (\mybeta_0\oplus\cdots\oplus\mybeta_{d-1})^\bot$, for any nonzero $v \in \mybeta_0 \oplus \mybeta_1 \oplus \cdots \oplus \mybeta_{d-1}$ one
has $\langle v,C_0 y \rangle = 0$ for all $y\in \myH$, and hence also for $y=s_{d,t}$ in \eqref{eq_CT}; this implies that $\langle v,x_t\rangle$ is at most $I(d-1)$, i.e. $\mybeta_0 \oplus \mybeta_1 \oplus \cdots \oplus \mybeta_{d-1}$ is the infinite dimensional \cs.
Note that this decomposition is orthogonal.

Using orthogonal projections, one can see that this orthogonal direct sum decomposition can be employed in general to characterize the degree of integration of any $v$-characteristic of the process.
In fact, note that \eqref{eq_Id_cond} implies $P_{\tau_0}+P_{\tau_1}+\dots+P_{\tau_d}=I$, where $P_{\tau_h}$ is the orthogonal projection onto $\tau_h$; hence for any nonzero $v\in \myH$ one has $\langle v,x_t\rangle=\langle v_0,x_t\rangle+\langle v_1,x_t\rangle+\dots+\langle v_d,x_t\rangle$, where $v_h=P_{\tau_h}v\in \tau_h$.
\eqref{eq_char_Id} describes the order of integration of any nonzero characteristic $\langle v,x_t\rangle$ of $x_t$. In particular one has $\langle v,x_t\rangle\sim I(d)$ if and only if $v_{d}\neq 0$, because $\langle v_0,x_t\rangle+\langle v_1,x_t\rangle+\dots+\langle v_{d-1},x_t\rangle$ is at most $I(d-1)$.
\end{remark}

\begin{remark}\label{rem-14}
Theorem \ref{thm_CHAR_d} further shows how the properties of $\langle v,x_t\rangle$ vary with $v\in \mybeta_0\oplus \mybeta_1\oplus\dots\oplus\mybeta_{d-1}$: in $\mybeta_0$, which is infinite dimensional, one finds the cointegrating vectors that allow for polynomial cointegration of order 0, i.e. for any nonzero $v\in \mybeta_0$, one has $\langle v,x_t\rangle + \sum_{n=1}^{d-1}\langle v,A_0^+A_{n}\Delta^n x_t\rangle \sim I(0)$, while in $\mybeta_h$, $h=1,\dots,d-2$, which is finite dimensional and can as well be equal to 0, those that allow for polynomial cointegration of order $h$, i.e. for any nonzero $v\in \mybeta_{1}$, one has $\langle v,x_t\rangle + \sum_{n=1}^{d-2}\langle v,S_1^+A_{2,n}\Delta^n x_t\rangle \sim I(1)$, for any nonzero $v\in \mybeta_{2}$, one has $\langle v,x_t\rangle + \sum_{n=1}^{d-3}\langle v,S_2^+A_{3,n}\Delta^n x_t\rangle \sim I(2)$ and so on up to nonzero $v\in \mybeta_{d-2}$, for which $\langle v,x_t\rangle + \langle v,S_{d-2}^+A_{d-1,1}\Delta x_t\rangle \sim I(d-2)$. In $\mybeta_{d-1}$, with $0\leq \dim \mybeta_{d-1}< \infty$, every nonzero linear combination of $x_t$ is $I(d-1)$ and does not allow for polynomial cointegration and in $\mybeta_{d}$, with $0 < \dim \mybeta_{d}< \infty$, every nonzero linear combination of $x_t$ is $I(d)$.
\end{remark}

As discussed in the next remark, the only difference with the finite dimensional case, see \citet{FP_idcoeff}, is that in that case the number of cointegrating relations of order 0 is finite.

\begin{remark}\label{rem_Id_finite}
In the finite dimensional case $\myH=\BR^p$, \citet{FP:16} show that $d=1,2,\dots$ if and only if $\BR^p=\myalpha_0 \oplus \cdots \oplus \myalpha_d=\mybeta_0 \oplus \cdots \oplus \mybeta_d$, where $\myalpha_{h}=\msp\{\alpha_h\}$, $\mybeta_{h}=\msp\{\beta_h\}$, $h=0,1,\dots$, and the bases $\alpha_h$, $\beta_h$ are defined by the \rf s $P_{\mya_{h}^\bot}A_{h,1}P_{\myb_{h}^\bot}=\alpha_h\beta_h'$, where 
$A_{h,1}$ is as in Definition \tref{def_LRF} with $S^+_h=\bar{\beta}_h\bar{\alpha}_h'$. Again here, apart from the fact that $\dim \myalpha_0=\dim \mybeta_0$ is finite when $\myH=\BR^p$, this mirrors what happens in the infinite dimensional case.
\end{remark}

\begin{remark}\label{rem_Id_iota}
The \pole{d} condition in \eqref{eq_Id_cond} is equivalent to $\mybeta_d=(\mybeta_0\oplus\dots\oplus\mybeta_{d-1})^\bot\neq \{ 0 \}$. Moreover, Theorem \tref{thm_pole_order} in Appendix \tref{app_lemmas} shows that it can be equivalently stated as $(i)$ $\myH=\myalpha_0 \oplus \myalpha_1 \oplus \cdots \oplus \myalpha_d$, $\myalpha_d\neq \{ 0 \}$, $(ii)$ $\myalpha_d=(\myalpha_0\oplus\dots\oplus\myalpha_{d-1})^\bot\neq \{ 0 \}$, $(iii)$ $\Ima C_0=\mybeta_d$, $(iv)$ $\Ker C_0=\mybeta_0\oplus\dots\oplus\mybeta_{d-1}$.
\end{remark}

In order to complete the discussion of the relation of the present results with the existing literature, the equivalence of the \pole{d} condition in \eqref{eq_Id_cond} to the condition in \citet{HP:WP} reported in eq. \eqref{eq_HP_Id} below is discussed.

\citet{HP:WP} consider $x_t=A_1^{\circ}x_{t-1}+\varepsilon_t$ with $A_1^{\circ}$ compact and formulate an $I(d)$ condition and then study the $I(1)$ case. In order to state their $I(d)$ condition, they employ the nonorthogonal direct sum decomposition $\myH=\myH_P \oplus \myH_T$, where $\myH_P$ is the finite dimensional image of the Riesz projection associated with the isolated eigenvalue $z=1$ and $\myH_T$ is the infinite dimensional image of the Riesz projections associated with the remaining stable eigenvalues. Using the nonorthogonal projections associated to the nonorthogonal direct sum decomposition $\myH=\myH_P \oplus \myH_T$, they decompose the process into $x_t=x_t^P+x_t^T$, where
$x_t^X= A_X x_t^X+ \varepsilon_t^X \in \myH_X$ and $A_X$ is the restriction of $A_1^{\circ}$ to $\myH_X$, $X=T,P$. Their $I(d)$ condition is stated as
\begin{equation}\label{eq_HP_Id}
A_P-I \mbox{ is a nilpotent matrix of order } d,
\end{equation}
i.e. $(A_P-I)^{d-1}\neq0$ and $(A_P-I)^{d}=0$, which simplifies to $A_P=I$ in the $I(1)$ case studied in that paper.

\begin{prop}\label{prop_equiv_HP}
Let $A(L)x_{t}=\varepsilon_{t}$ be an \ar; then the $I(d)$ condition in \eqref{eq_HP_Id} is equivalent to the \pole{d} condition in \eqref{eq_Id_cond}.
\end{prop}

\section{Conclusion}\label{sec_CONC}

The present paper characterizes the cointegration properties of \ars, i.e. $\myH$-valued AR processes $A(L)x_{t}=\varepsilon_{t}$ such that $A(z)$ has an eigenvalue of finite type at $z=1$ and it is invertible in the punctured disc $D(0,\rho)\setminus\{1\}$ for some $\rho>1$. It is shown that \ars\ are necessarily integrated of finite order $d$ and necessarily have a finite number of $I(d)$ trends and an infinite dimensional \cs. This is in line with the setup employed in most contributions in the literature and seems to be the most empirically relevant framework.

A necessary and sufficient condition on the AR operators that establishes the value of $d$ is given in terms of the orthogonal direct sum decomposition $\myH=\mybeta_0\oplus\mybeta_1\oplus\dots\oplus\mybeta_d$, $\mybeta_d\neq \{ 0 \}$, where $\mybeta_0$ is infinite dimensional, $0 \leq \dim \mybeta_h < \infty$, $h=1,\dots,d-1$, with strict inequality for $h = d$.

A full description of how the properties of the characteristic $\langle v,x_t\rangle$ vary with $v\in \myH$ is given: in $\mybeta_0$, one can combine $\langle v,x_t\rangle$ with differences of the process and find at most $I(0)$ polynomial cointegrating relations, in $ \mybeta_1$, one can combine $\langle v,x_t\rangle$ with differences and find at most $I(1)$ polynomial cointegrating relations, and so on up to $\mybeta_{d-2}$, in which one can combine $\langle v,x_t\rangle$ with differences and find at most $I(d-2)$ polynomial cointegrating relations. Finally, any nonzero $v\in\mybeta_{d-1}$ is such that $\langle v,x_t\rangle$ is $I(d-1)$ and does not allow for polynomial cointegration and any nonzero $ v\in \mybeta_{d}$ is such that $\langle v,x_t\rangle$ is $I(d)$. This shows that the infinite dimensional subspace $\mybeta_0\oplus\mybeta_1\oplus\dots\oplus\mybeta_{d-1}$ is the \cs\ while the finite dimensional subspace $\mybeta_d$ is the \as.

For any nonzero $v$ in the \cs, the expression of the polynomial cointegrating relations is provided in terms of operators that are defined recursively in terms of the AR operators together with the $\mybeta_h$.

The present results show that, under the assumption that 1 is an eigenvalue of finite type of the AR operator function, the infinite dimensionality of the space does not introduce additional elements in the analysis. That is, apart from the fact that the number of $I(0)$ cointegrating relations is infinite, conditions and properties of $\myH$-valued AR processes coincide with those that apply in the usual finite dimensional VAR case.

\cp

\appendix

\section{Notation and background results}\label{app_dot}

In the present paper $\myH$ is an infinite dimensional separable Hilbert space and a random variable that takes values in $\myH$ is said to be an $\myH$-valued random variable and a sequence of $\myH$-valued random variables is called an $\myH$-valued stochastic process. Section \tref{app_operator} reviews notions and results on separable Hilbert spaces and on operators acting on them and Section \tref{app_random} presents the definitions of expectation and covariance operator for $\myH$-valued random variables.

\subsection{Separable Hilbert spaces and operators acting on them}\label{app_operator}
The material in this section is based on Chapters I, II in \citet{GGK:03} and Chapter XI in \citet{GGK:90:1}.
Let $\myH$ be a separable Hilbert space with inner product $\langle \: \cdot \:,\cdot \:\rangle$ and norm $\| x \|=\langle x,x\rangle ^{\frac{1}{2}}$; a function $A:\myH \rightarrow \myH$, is called a linear operator if for all $v,w \in \myH$ and $c\in\BC$, $A(v+w)=Av+A w$ and $A(c v) = c Av$, where $A u$ and $A[u]$ both indicate the action of $A$ on $u\in \myH$. A linear operator $A$ is called bounded if its norm $\| A \|_{\myB}=\sup_{\| v \|=1}\| Av \|$ is finite and the set of bounded linear operators with norm $\|\cdot\|_{\myB}$ is denoted as $\myB$. For any $A \in\myB$ the subspace $\{v\in \myH:Av=0 \}$, written $\Ker A$, is called the kernel of $A$ and the subspace $\{Av : v\in \myH\}$, written $\Ima A$, is called the image of $A$. The dimension of $\Ima A$, written $\dim \Ima A$, is called the rank of $A$, written $\rank A$.

$\myH$ is said to be the direct sum of subspaces $S$ and $U$, written $\myH=S\oplus U$, if $S\cap U=0$ and every vector $v\in \myH$ can be written as $v=s+u$, where $s\in S$ and $u\in U$. The set $\{v\in \myH: \langle v,s\rangle=0 \mbox{ for all }s\in S \subseteq \myH\}$ is called the orthogonal complement of $S$, written $S^\bot$. For $U=S^\bot$, one has the orthogonal direct sum $\myH=S\oplus S^\bot$. The orthogonal projection on $\eta$, written $P_\eta$, is such that $P_\eta\in\myB$, $P_\eta^2=P_\eta$, $\Ima P_\eta=\eta$ and $\Ker P_\eta=\eta^\bot$; moreover, $I=P_{\eta}+P_{\eta^\bot}$.

An operator $A\in\myB$ is said to be invertible if there exists an operator $B\in \myB$ such that $BAv=ABv=v$ for every $v\in \myH$; in this case $B$ is called the inverse of $A$, written $A^{-1}$.
An operator $A\in\myB$ such that $n(A) = \dim \Ker A<\infty$ and $d(A) = \dim (\Ima A)^\bot<\infty$ is said to be Fredholm of index $n(A)-d(A)$. Remark that if $\myH$ is finite dimensional, any $A\in\myB$ is \fof.

Corollary 8.4 in Section XI.8 in \citet{GGK:90:1} states that the inverse of an operator function that is \fof\ and non-invertible at some isolated point has a pole at that point. Moreover, the operators that make up the principal part of its Laurent representation around that point have finite rank. If $z_0$ is an eigenvalue of finite type of $W(z)$, see Definition \tref{def_eig_finite_type}, then $z_0$ is an isolated singularity of $W(z)^{-1}$, $W(z_0)$ is \fof\ and non-invertible at $z_0$, so that Theorem \tref{thm_FLSF} below applies.

\begin{thm}\label{thm_FLSF}
Let $z_0$ be an eigenvalue of finite type of an operator function $W(z)$. Then there exist a finite integer $d=1,2,\dots$ and finite rank operators $U_{0},U_{1},\dots,U_{d-1}$ such that
$$
W(z)^{-1}=\sum_{n=0}^{\infty} U_n (z-z_0)^{n-d}, \qquad z\in D(z_0,\delta)\setminus\{z_0\},
$$
where $U_{d}$ is \fof.
\end{thm}
\begin{mproof}{\textit{Proof}.}
See Section XI.9 in \citet{GGK:90:1}.
\end{mproof}

\subsection{Random variables in separable Hilbert spaces}\label{app_random}
The definitions in this section are taken from Chapter 1 in \citet{Bos:00}.
Let $\myH$ be a separable Hilbert space with inner product $\langle \: \cdot \:,\cdot \:\rangle$, norm $\| w \| = \langle w ,w \rangle ^{\frac{1}{2}}$, and Borel $\sigma$-algebra $\sigma(\myH)$ and let $(\Omega,\mathcal{A},P)$ be a probability space. A function $Z:\Omega \rightarrow \myH$ is called an $\myH$-valued random variable on $(\Omega,\mathcal{A},P)$ if it is measurable, i.e. for every subset $S\in \sigma(\myH)$, $\{\omega :Z(\omega)\in S\}\in \mathcal {A}$. For a $\BC$-valued random variable $X$ on $(\Omega,\mathcal{A},P)$, define $\E(X)=\int_\Omega X(\omega) dP(\omega)$; the expectation of an $\myH$-valued random variable $Z$, written $\E(Z)$, is defined as the unique element $\mu$ of $\myH$ such that
$$
\E(\langle v,Z \rangle)=\langle v,\mu \rangle \mbox{ for all } v\in \myH.
$$
It can be shown that the existence of $\E(Z)$ is guaranteed by the condition $\E(\|Z\| )<\infty$. The covariance function of an $\myH$-valued random variable $Z$ is defined as
$$
c_Z(v,w)=\E(\langle v,Z-\E(Z) \rangle \langle w,Z-\E(Z) \rangle),\qquad v,w\in \myH.
$$
It is immediate to see that $c_Z(v,w)=
\E(\langle v,W \rangle)-\langle v,\E(Z) \rangle \langle w,\E(Z) \rangle$, where $W=\langle w,Z \rangle Z$. If $\E(\|W\|)<\infty$, the expectation of the $\myH$-valued random variable $W$ exists and it is the unique element of $\myH$ such that $\E(\langle v,W \rangle)=\langle v,\E(W) \rangle$ for all $v\in \myH$. One thus has
$$
c_Z(v,w)=\langle v,\E(W) \rangle-\langle v,\E(Z) \rangle \langle w,\E(Z) \rangle,\qquad v,w\in \myH, \qquad W=\langle w,Z \rangle Z.
$$
Because $\|W\|=|\langle w,Z \rangle|\| Z\|\leq \| w \|\| Z\|^2$, the existence of the covariance function of $Z$ is guaranteed by the condition $\E(\|Z\|^2)<\infty$. Define the operator $C_Z:\myH\rightarrow \myH$ that maps $w$ into $\E(W)$ and rewrite the covariance function as
$c_Z(v,w)=\langle v,C_Zw \rangle-\langle v,\E(Z) \rangle \langle w,\E(Z) \rangle,\, v,w\in \myH.$
$C_Z$ is fully determined by the covariance function and it is called the covariance operator of $Z$.
Similarly, the cross-covariance function of two $\myH$-valued random variables $Z$ and $U$ is defined as
$$
c_{Z,U}(v,w)=E(\langle v,Z-\E(Z) \rangle \langle w,U-\E(U) \rangle),\qquad v,w\in \myH.
$$
This also completely determines the cross-covariance operators of $Z$ and $U$, $C_{Z,U}$ and $C_{U,Z}$, respectively defined as the mappings $w\mapsto \E(\langle w,Z \rangle U)$ and $w\mapsto \E (\langle w,U \rangle Z)$.

\section{Inversion of an operator function around a singular point}\label{app_lemmas}

This Appendix presents novel results on the inversion of a meromorphic operator function which are used in Appendix \tref{app_text} to prove the results in the text.

The inversion results are derived from system \eqref{eq_IDENT_RAW} below, see e.g. \citet{HAPE:09}. When the inverse $A(z)^{-1}$ has a pole of order $d$ from the identity $A(z)A(z)^{-1}=I=A(z)^{-1}A(z)$ one finds the following linear systems in the $A_n$, $C_n$ operators defined in \eqref{eq_def_Az} and \eqref{eq_LAU2},
\begin{align}
A_{0}C_{0}= & \: 0=C_{0}A_{0} \notag \\
A_{0}C_{1}+A_{1}C_{0}= & \: 0=C_{0}A_{1}+C_{1}A_{0} \notag \\
& \: \vdots \notag \\
A_{0}C_{d-1}+\dots +A_{d-1}C_{0}= & \: 0=C_{0}A_{d-1}+\dots +C_{d-1}A_{0} \label{eq_IDENT_RAW} \\
A_{0}C_{d}+A_{1}C_{d-1}+\dots +A_{d}C_{0}= & \: I=C_{0}A_{d}+C_{1}A_{d-1}+\dots +C_{d}A_{0} \notag \\
A_{0}C_{d+1}+A_{1}C_{d}+\dots +A_{d+1}C_{0}= & \: 0=C_{0}A_{d+1}+C_{1}A_{d}+\dots +C_{d+1}A_{0} \notag \\
& \: \vdots \notag
\end{align}
In the following, equations in system \eqref{eq_IDENT_RAW} are numbered according to the highest value of the subscript of $C_{n}$. Note that the identity appears in equation $d$, which is the order of the pole. The equations that derive from $A(z)A(z)^{-1}=I$ are called left versions (and correspond to the left side of \eqref{eq_IDENT_RAW}) and those that derive from $I=A(z)^{-1} A(z)$ are called right versions (and correspond to the right side of \eqref{eq_IDENT_RAW}). For instance $A_{0}C_{d}+A_{1}C_{d-1}+\dots +A_{d}C_{0}=I$ is called the left version of equation $d$.

Recall that $P_\eta\in \myB$ indicates the orthogonal projection on $\eta$ 
and $A^+$ and $A^*$ respectively denote the generalized inverse and the adjoint of $A$.

\begin{lemma}\label{lem_gen_inv_h}
Consider Definition \tref{def_LRF}. Then $\Ker S^+_h =(\Ima S_h)^\bot$, $S^+_h S_h=P_{\mybeta_h}$ and $S^+_h P_{\mya_{h}^\bot}=S^+_h$, $h=0,1,\dots,d$.
\end{lemma}
\begin{mproof}{\textit{Proof}.}
From Theorem 3 in \citet[][Chapter 9]{BI:03}, one has $S^+_hS_h=P_{\Ima S_h^*}$ and $\Ker S^+_h = \Ker S_h^*$ and from Theorem 11.4 in \citet[][Chapter II]{GGK:03} one has $\Ima S_h^*=(\Ker S_h)^\bot$ and $\Ker S_h^*=(\Ima S_h)^\bot$, so that $\Ker S^+_h =(\Ima S_h)^\bot$. By Definition \tref{def_LRF}, $(\Ker S_h)^\bot=\mybeta_h$ and hence $S^+_h S_h=P_{\mybeta_h}$. Moreover, by Definition \tref{def_LRF}, $(\Ima S_h)^\bot=\myalpha_h^\bot\supseteq \myalpha_0 \oplus \cdots \oplus \myalpha_{h-1}=\mya_{h}$ and hence $\mya_{h}\subseteq \Ker S^+_h$, which implies $S^+_h=S^+_h P_{\mya_{h}^\bot}$.
\end{mproof}

\begin{lemma}[Subspace decompositions of system \eqref{eq_IDENT_RAW}]\label{lem_LRF}
Consider Definition \tref{def_LRF} and further define $P_{\mya_{0}^\bot}=P_{\myb_{0}^\bot}=I$. Then the left version of equation $n+h\leq d$ in system \eqref{eq_IDENT_RAW} implies
\begin{equation}\label{eq_lrf_left}
S_h C_{n}+P_{\mya_{h}^\bot}\sum_{k=1}^{n}A_{h+1,k}C_{n-k}=\delta_{n+h,d} P_{\mya_{h}^\bot},\qquad h=0,1,\dots,d-n,
\end{equation}
where $\delta_{hj}$ is the Kronecker delta. Similarly, the right version of equation $n+h\leq d$ in system \eqref{eq_IDENT_RAW} implies
\begin{equation}\label{eq_lrf_right}
C_{n}S_h+\sum_{k=1}^{n}C_{n-k}A_{h+1,k}P_{\myb_{h}^\bot}=\delta_{n+h,d} P_{\myb_{h}^\bot},\qquad h=0,1,\dots,d-n.
\end{equation}
\end{lemma}
\begin{mproof}{\textit{Proof}.}
The proof of \eqref{eq_lrf_left} is by induction and consists in showing that the left version of equation $n\leq d$ in system \eqref{eq_IDENT_RAW} implies
\begin{equation}\label{eq_lrf_pf}
S_h C_{n-h}+P_{\mya_{h}^\bot}\sum_{k=1}^{n-h}A_{h+1,k}C_{n-h-k}=\delta_{n,d} P_{\mya_{h}^\bot},\qquad h=0,1,\dots,n;
\end{equation}
replacing $n$ with $n+h$ one finds \eqref{eq_lrf_left}. In order to show that \eqref{eq_lrf_pf} holds for $h=0$, observe that the left version of equation $n$ in system \eqref{eq_IDENT_RAW} reads $A_{0}C_{n}+\sum_{k=1}^{n}A_{k}C_{n-k}=\delta_{n,d}I$. By definition, $P_{\mya_{0}^\bot}=I$, $S_0=A_0$ and $A_{1,k}=A_k$ and this shows that \eqref{eq_lrf_pf} holds for $h=0$. Next assume that \eqref{eq_lrf_pf} holds for $h=0,\dots,\ell-1$ for some $1<\ell\leq d$; one wishes to show that it also holds for $h=\ell$. First note that $S^+_h S_h=P_{\mybeta_h}$ and $S^+_h P_{\mya_{h}^\bot}=S^+_h$, see Lemma \tref{lem_gen_inv_h}; thus the induction assumption implies
$$
P_{\mybeta_h} C_{n-h}+S^+_h\sum_{k=1}^{n-h}A_{h+1,k}C_{n-h-k}=\delta_{n,d} S^+_h ,\qquad h=0,1,\dots,\ell-1,
$$
and replacing $n$ with $n-\ell+h$ and $h$ with $i$, one has
$$
P_{\mybeta_i} C_{n-\ell}=-S^+_i\sum_{k=1}^{n-\ell}A_{i+1,k}C_{n-\ell-k}+\delta_{n-\ell+i,d} S^+_i ,\qquad i=0,1,\dots,\ell-1.
$$
Observe that for $i=0,1,\dots,\ell-1$ one has $n-\ell+i\leq n-1 <d$; hence $\delta_{n-\ell+i,d}=0$ and one finds
\begin{equation}\label{eq_ind_ass}
P_{\mybeta_i} C_{n-\ell}=-S^+_i\sum_{k=1}^{n-\ell}A_{i+1,k}C_{n-\ell-k},\qquad i=0,1,\dots,\ell-1.
\end{equation}

Next write \eqref{eq_lrf_pf} for $h=\ell-1$,
$$
S_{\ell-1} C_{n-\ell+1}+P_{\mya_{\ell-1}^\bot}\sum_{k=1}^{n-\ell+1}A_{\ell,k}C_{n-\ell+1-k}=\delta_{n,d} P_{\mya_{\ell-1}^\bot},
$$
where $\Ima S_{\ell-1}=\myalpha_{\ell-1}$, see Definition \tref{def_LRF}; applying $P_{\mya_{\ell}^\bot}$, where $\mya_{\ell}=\myalpha_{0}\oplus\dots\oplus\myalpha_{\ell-1}$, one has $P_{\mya_{\ell}^\bot}S_{\ell-1}=0$ and rearranging one finds
\begin{equation}\label{eq_UV}
P_{\mya_{\ell}^\bot}A_{\ell,1}C_{n-\ell}+P_{\mya_{\ell}^\bot}\sum_{k=1}^{n-\ell}A_{\ell,k+1}C_{n-\ell-k}=\delta_{n,d} P_{\mya_{\ell}^\bot}.
\end{equation}
Next consider $\myb_{\ell}=\mybeta_{0}\oplus\dots\oplus\mybeta_{\ell-1}$ and use projections, inserting $I=P_{\myb_{\ell}^\bot}+P_{\myb_{\ell}}$ between $A_{\ell,1}$ and $C_{n-\ell}$ in $P_{\mya_{\ell}^\bot}A_{\ell,1}C_{n-\ell}=U$, say; one finds
$$
U=\left(P_{\mya_{\ell}^\bot}A_{{\ell},1}P_{\myb_{\ell}^\bot}\right)C_{n-\ell}+P_{\mya_{\ell}^\bot}A_{\ell,1}P_{\myb_{\ell}}C_{n-\ell}=U_{1}+U_{2},\mbox{ say}.
$$
By Definition \tref{def_LRF}, $P_{\mya_{\ell}^\bot}A_{{\ell},1}P_{\myb_{\ell}^\bot}=S_\ell$, so that $U=S_\ell C_{n-\ell}+U_2$. Substituting $P_{\myb_{\ell}}=P_{\mybeta_{0}}+\dots+P_{\mybeta_{\ell-1}}$ in $U_{2}$, one has $U_{2}= P_{\mya_{\ell}^\bot}A_{\ell,1}\sum_{i=0}^{\ell-1}P_{\mybeta_{i}}C_{n-\ell}$ and by the induction assumption, see \eqref{eq_ind_ass}, one finds
$$
U_2 = - P_{\mya_{\ell}^\bot}\sum_{k=1}^{n-\ell}\left(A_{\ell,1}\sum_{i=0}^{\ell-1}S^+_iA_{i+1,k}\right)C_{n-\ell-k}.
$$
Substituting the expression of $U_2$ into $U=S_\ell C_{n-\ell}+U_2$ and using $A_{\ell+1,k}=A_{\ell,k+1}-A_{\ell,1}\sum_{i=0}^{\ell-1}S^+_iA_{i+1,k}$, see Definition \tref{def_LRF}, one hence rewrites \eqref{eq_UV} as
$$
S_\ell C_{n-\ell}+P_{\mya_{\ell}^\bot}\sum_{k=1}^{n-\ell}A_{\ell+1,k}C_{n-\ell-k}=\delta_{n,d} P_{\mya_{\ell}^\bot}.
$$
This shows that \eqref{eq_lrf_pf} holds for $h=\ell$ and completes the proof of \eqref{eq_lrf_left}. A similar induction on the right version of system \eqref{eq_IDENT_RAW} leads to \eqref{eq_lrf_right}.
\end{mproof}

\begin{lemma}\label{lem_B0}
Consider Definition \tref{def_LRF}. Then $\Ima C_{0}\subseteq \myb_d^\bot$ and $\mya_d \subseteq \Ker C_{0}$.
\end{lemma}
\begin{mproof}{\textit{Proof}.}
For $n=0$, \eqref{eq_lrf_left} and \eqref{eq_lrf_right} read
\begin{equation}\label{eq_lrf_lr+n=0}
S_h C_{0}=\delta_{h,d} P_{\mya_{h}^\bot},\qquad C_{0}S_h=\delta_{h,d} P_{\myb_{h}^\bot}\qquad h=0,1,\dots,d,
\end{equation}
where $S_h=P_{\mya_{h}^\bot}A_{{h},1}P_{\myb_{h}^\bot}$, see Definition \tref{def_LRF}. \eqref{eq_lrf_lr+n=0} implies $S_h C_{0}=C_{0}S_h=0$ for $h=0,1,\dots,d-1$. From $S_h C_{0}=0$, $h=0,1,\dots,d-1$, one has  $\Ima C_0 \subseteq \Ker S_h$ for $h=0,1,\dots,d-1$, i.e. $\Ima C_0 \subseteq \left(\Ker S_0\cap \Ker S_1 \cap \dots \cap \Ker S_{d-1}\right)$. By Definition \tref{def_LRF}, $\Ker S_h=\mybeta_h^\bot$ and hence $\Ima C_0 \subseteq \left(\mybeta_0^\bot\cap \mybeta_1^\bot \cap \dots \cap \mybeta_{d-1}^\bot\right)=(\mybeta_0\oplus \mybeta_1 \oplus \dots \oplus \mybeta_{d-1})^\bot=\myb_d^\bot$. This proves the first statement.

From $C_{0}S_h =0$, $h=0,1,\dots,d-1$, one has  $\Ima S_h \subseteq \Ker C_0$ for $h=0,1,\dots,d-1$, i.e. $\left(\Ima S_0\oplus \Ima S_1 \oplus \dots \oplus \Ima S_{d-1}\right) \subseteq \Ker C_0$. By Definition \tref{def_LRF}, $\Ima S_h=\myalpha_h$ and hence $ \myalpha_0\oplus \myalpha_1 \oplus \dots \oplus \myalpha_{d-1} =\mya_d \subseteq \Ker C_0$.
\end{mproof}

\begin{thm}[Order of the pole]\label{thm_pole_order}
Consider Definition \tref{def_LRF}. The following statements are equivalent:

$(i)$ $A(z)^{-1}$ has a pole of order $d$ at $z=1$,

$(ii)$ the identity is in equation $d$ of system \eqref{eq_IDENT_RAW},

$(iii)$ $\myalpha_d=(\myalpha_0 \oplus \cdots \oplus \myalpha_{d-1})^\bot\neq \{ 0 \}$,

$(iv)$ $\Ker C_0=\myalpha_d^\bot$,

$(v)$ $\mybeta_d=(\mybeta_0 \oplus \cdots \oplus \mybeta_{d-1})^\bot\neq \{ 0 \}$,

$(vi)$ $\Ima C_0=\mybeta_d$.
\end{thm}
\begin{mproof}{\textit{Proof}.}

\noindent$(i) \miff (ii)$ By definition.

\noindent $(ii) \rimp (iii) \rimp (iv)$. Under $(ii)$, one has $h=d$ in the left equation in \eqref{eq_lrf_lr+n=0}, i.e. $S_d C_{0}=P_{\mya_{d}^\bot}$, $\mya_{d}^\bot\neq \{ 0 \}$; by Definition \tref{def_LRF}, $\Ima S_d \subseteq \mya_d^\bot$ and because $\Ima S_d \subset \mya_d^\bot$ contradicts $S_d C_{0}=P_{\mya_{d}^\bot}$, one has  $\Ima S_d =\mya_{d}^\bot$. By Definition \tref{def_LRF}, $\Ima S_d = \myalpha_d$ and $\mya_{d}^\bot= (\myalpha_0 \oplus \cdots \oplus \myalpha_{d-1})^\bot$, and hence $(iii)$. Moreover, by Lemma \tref{lem_B0}, $\mya_d \subseteq \Ker C_{0}$ and because $\mya_d \subset \Ker C_{0}$ contradicts $S_d C_{0}=P_{\mya_{d}^\bot}$, one has  $\mya_d = \Ker C_{0}$. Using $\mya_d=\myalpha_d^\bot$, see $(iii)$, one finds $(iv)$.

\noindent $(iv) \rimp (ii)$. Let $\Ker C_0=\myalpha_d^\bot$ and proceed by contradiction, assuming that the identity is not in equation $d$, so that the right equation in \eqref{eq_lrf_lr+n=0} reads $C_{0}S_d=0$, which implies $\Ima S_d \subseteq \Ker C_0$, where $\Ima S_d=\myalpha_d$ and $\Ker C_0=\myalpha_d^\bot$. Hence $\myalpha_d\subseteq\myalpha_d^\bot$, so that $\myalpha_d=\{0\}$ and thus $\myalpha_d^\bot=\myH$. This contradicts $C_0\neq 0$, i.e. that the pole has order $d$, and proves that $(ii)$ holds.

\noindent $(ii) \rimp (v) \rimp (vi)$. Under $(ii)$, one has $h=d$ in the right equation in \eqref{eq_lrf_lr+n=0}, i.e. $C_{0}S_d =P_{\myb_{d}^\bot}$, $\myb_{d}^\bot\neq \{ 0 \}$; by Definition \tref{def_LRF}, $\myb_d \subseteq \Ker S_d$ and because $\myb_d \subset \Ker S_d$ contradicts $C_{0} S_d =P_{\myb_{d}^\bot}$, one has  $\myb_d = \Ker S_d$. By Definition \tref{def_LRF}, $\Ker S_d = \mybeta_d^\bot$ and $\myb_{d}= \mybeta_0 \oplus \cdots \oplus \mybeta_{d-1}$, and hence $(v)$. Moreover, by Lemma \tref{lem_B0}, $\Ima C_{0} \subseteq \myb_d^\bot$ and because $\Ima C_{0} \subset \myb_d^\bot$ contradicts $C_{0} S_d =P_{\myb_{d}^\bot}$, one has  $\Ima C_{0} = \myb_d^\bot$. Using $\myb_d^\bot=\mybeta_d$, see $(v)$, one finds $(vi)$.

\noindent $(vi) \rimp (ii)$. Let $\Ima C_0=\mybeta_d$ and proceed by contradiction, assuming that the identity is not in equation $d$, so that the left equation in \eqref{eq_lrf_lr+n=0} reads $S_d C_{0}=0$, which implies $\Ima C_0 \subseteq \Ker S_d$, where $\Ima C_0=\mybeta_d$ and $\Ker S_d=\mybeta_d^\bot$. Hence $\mybeta_d\subseteq\mybeta_d^\bot$, so that $\mybeta_d=\{0\}$. This contradicts $C_0\neq 0$, i.e. that the pole has order $d$, and proves that $(ii)$ holds.
\end{mproof}

\begin{thm}[Pole cancellations in $A(z)^{-1}$]\label{thm_pole_canc}
Consider Definition \tref{def_LRF} and for $h=0,1,\dots,d$ define
$$
\gamma_{h}(z)=P_{\mybeta_h}+S^+_h\sum_{n=1}^{d-h-1} A_{h+1,n}(1-z)^n.
$$
Then $\langle v,\gamma_{h}(z) A(z)^{-1} y \rangle$ has a pole of order $h=0,1,\dots,d$ for any nonzero $v\in \mybeta_h$ and $y \in \myH$.
\end{thm}
\begin{mproof}{\textit{Proof}.}
Applying $S^+_h$ to \eqref{eq_lrf_left} and using $S^+_h S_h=P_{\mybeta_h}$ and $S^+_h P_{\mya_{h}^\bot}=S^+_h$, see Lemma \tref{lem_gen_inv_h}, one finds
\begin{equation}\label{eq_gamma}
P_{\mybeta_{h}}C_{n}+S^+_h\sum_{k=1}^{n}A_{h+1,k}C_{n-k}=\delta_{n+h,d} S^+_h,\qquad h=0,1,\dots,d-n.
\end{equation}
Write $A(z)^{-1}=\sum_{n=0}^{\infty}C_{n}(1-z)^{n-d}$ as
$$
A(z)^{-1}=C_{0}(1-z)^{-d}+\sum_{n=1}^{d-h-1}C_{n}(1-z)^{n-d}+(1-z)^{-h}R_0(z),\qquad R_0(1) = C_{d-h},
$$
and apply $P_{\mybeta_h}$ to find
$$
P_{\mybeta_h}A(z)^{-1}=P_{\mybeta_h}C_{0}(1-z)^{-d}+\sum_{n=1}^{d-h-1}P_{\mybeta_h}C_{n}(1-z)^{n-d}+(1-z)^{-h}P_{\mybeta_h}R_0(z).
$$
First consider $h=0,\dots,d-1$. Setting $n=0$ in \eqref{eq_gamma} one has $P_{\mybeta_{h}}C_{0}=0$ and hence
\begin{equation}\label{eq_kappa_h_Az_inv}
P_{\mybeta_h}A(z)^{-1}=\sum_{n=1}^{d-h-1}P_{\mybeta_h}C_{n}(1-z)^{n-d}+(1-z)^{-h}P_{\mybeta_h}R_0(z).
\end{equation}
From \eqref{eq_gamma}, for $n\leq d-h$ one has $P_{\mybeta_{h}}C_{n}=-S^+_h\sum_{k=1}^{n}A_{h+1,k}C_{n-k}+\delta_{n+h,d} S^+_h$ and because $\delta_{n+h,d}=0$ for $n=1,\dots, d-h-1$, one has
$$
\sum_{n=1}^{d-h-1}P_{\mybeta_h}C_{n}(1-z)^{n-d}=-\sum_{n=1}^{d-h-1}\left(S^+_h\sum_{k=1}^{n}A_{h+1,k}C_{n-k}\right)(1-z)^{n-d}.
$$
Rearraging one thus finds
$$
\sum_{n=1}^{d-h-1}P_{\mybeta_h}C_{n}(1-z)^{n-d}=-S^+_h\sum_{k=1}^{d-h-1}A_{h+1,k}\left( \sum_{n=k}^{d-h-1}C_{n-k}(1-z)^{n-d} \right).
$$
Next write
$$
(1-z)^k A(z)^{-1}=\left(\sum_{n=k}^{d-h-1}C_{n-k}(1-z)^{n-d}\right)+(1-z)^{-h}R_{k}(z),\qquad R_k(1)=C_{d-h-k},
$$
so that
$$
\sum_{n=1}^{d-h-1}P_{\mybeta_h}C_{n}(1-z)^{n-d}=-\left(S^+_h\sum_{k=1}^{d-h-1} A_{h+1,k}(1-z)^k \right) A(z)^{-1}+(1-z)^{-h}S^+_h\sum_{k=1}^{d-h-1} A_{h+1,k}R_{k}(z).
$$
Substituting in \eqref{eq_kappa_h_Az_inv} and rearraging one thus finds
$$
\gamma_{h}(z) A(z)^{-1}=(1-z)^{-h}\wt{\gamma}_{h}(z),
$$
where
$$
\gamma_{h}(z)=P_{\mybeta_h}+S^+_h\sum_{k=1}^{d-h-1} A_{h+1,k}(1-z)^k ,\qquad \wt{\gamma}_{h}(z)=P_{\mybeta_h}R_0(z)+S^+_h\sum_{k=1}^{d-h-1} A_{h+1,k}R_{k}(z).
$$

Note that, because $R_k(1)=C_{d-h-k}$, one has
$$
\wt{\gamma}_{h}(1)= P_{\mybeta_h}C_{d-h}+S^+_h\sum_{k=1}^{d-h-1} A_{h+1,k}C_{d-h-k};
$$
from \eqref{eq_gamma} for $n=d-h$ one finds $P_{\mybeta_{h}}C_{d-h}+S^+_h\sum_{k=1}^{d-h}A_{h+1,k}C_{d-h-k}=S^+_h$, so that $\wt{\gamma}_{h}(1)=S^+_h(I-A_{h+1,d-h}C_{0})$. Using $S^+_hS_{h}=P_{\mybeta_h}$ and $C_0S_h=0$ one finds $\wt{\gamma}_{h}(1)S_h=P_{\mybeta_h}$. This shows that $\langle v,\wt{\gamma}_{h}(1)y\rangle\neq 0$ for any nonzero $v\in \mybeta_h$ and any nonzero $y\in \myH$ and hence $\langle v,\gamma_{h}(z) A(z)^{-1} y \rangle$ has a pole of order $h$ for any nonzero $v\in \mybeta_h$, $h=0,\dots,d-1$, and any nonzero $y\in \myH$. Finally consider $h=d$. Setting $n=0$ and $h=d$ in \eqref{eq_gamma} one has $P_{\mybeta_{d}}C_{0}=S^+_d$ and using $S^+_dS_{d}=P_{\mybeta_d}$ one finds $P_{\mybeta_{d}}C_{0}S_d=P_{\mybeta_d}$; this shows that $\langle v,C_{0} y \rangle\neq  0 $ for any nonzero $v\in \mybeta_d$ and any nonzero $y\in\myH$. Hence $\langle v,A(z)^{-1}y\rangle$ has a pole of order $d$ for any nonzero $v\in \mybeta_d$ and any nonzero $y\in\myH$. This completes the proof.
\end{mproof}

\section{Proofs}\label{app_text}

\hfill

This Appendix contains the proofs of the results in the text. The proof Theorem \tref{thm_EXI} makes use of the following fact, which is proven in \citet{FP_idcoeff}: for $t\in\BZ$, one has
\begin{equation}\label{eq_S_D_h_leq_s}
\ops^{s}\Delta^{h}u_{t} =\ops^{s-h}u_{t}-\sum_{n=s-h}^{s-1}\varsigma_{n,t}\Delta^{h-s+n}u_{0},\qquad 0<h\leq s,
\end{equation}
where $\varsigma_{n,t}$ is a polynomial of order $n$ in $t$.

\smallskip

\begin{mproof}{\textit{Proof of Theorem \tref{thm_EXI}}.}
The result is a direct consequence of Theorem \tref{thm_FLSF} in Appendix \tref{app_operator}. By definition, an \ar\ $A(L)x_{t}=\varepsilon_{t}$ is such that $A(1)\neq 0$, $A(z)$ has an eigenvalue of finite type at $z=1$ and $A(z)$ is invertible in the punctured disc $D(0,\rho)\setminus\{1\}$ for some $\rho>1$. Letting $z_0=1$ and $C_n=U_n (-1)^{n-d}$, Theorem \tref{thm_FLSF} states that there exist a finite integer $d=1,2,\dots$ and finite rank operators $C_{0},C_{1},\dots,C_{d-1}$ such that
\begin{equation}\label{eq_LAU}
A(z)^{-1}=\sum_{n=0}^{\infty} C_n (1-z)^{n-d}, \qquad z\in D(1,\delta)\setminus\{1\}.
\end{equation}
Write $A(z)^{-1}=\sum_{n=0}^{d-1} C_n (1-z)^{n-d}+C_d^\star(z)$, where $C_d^{\star}(z)$ is absolutely convergent in $D(0,\rho)$ for some $\rho>1$; applying $A(L)^{-1}$ on both sides of $A(L)x_t=\varepsilon_t$ one finds the common trends representation $x_t=C_{0}s_{d,t}+C_{1}s_{d-1,t}+\dots+C_{d-1}s_{1,t}+y_{t}+\mu_{t}$,
where $s_{h,t}=\ops ^h \varepsilon_{t}\sim I(h)$ is the $h$-fold integrated bilateral random walk, $y_{t}=C_d^{\star}(L)\varepsilon_{t}$ is a linear process, $\mu_t=\sum_{n=0}^{d-1}v_{n}t^n$, $v_n\in\myH$, is a polynomial of time, with coefficients $v_0,\dots,v_{d-1} \in \myH$, depend on the initial values of $x_{t},y_{t},\varepsilon_{t}$ for $t=-d,\dots,0$, see \eqref{eq_S_D_h_leq_s}.
\end{mproof}

\smallskip

\begin{mproof}{\textit{Proof of Corollary \tref{coro_EXI_CI}}.}
The order $d$ of the pole of the inverse of $A(z)$ is finite by Theorem \tref{thm_FLSF} in Appendix \tref{app_operator}; this implies $x_{t}\sim I(d)$ via \eqref{eq_CT}. The $I(d)$ trends $s_{d,t}$ are loaded onto $x_t$ by $C_0$, which has finite rank, implication $(iii)$, and hence any nonzero $v \in (\Ima C_0)^\bot$ is such that $\langle v,C_0 y \rangle = 0$ for all $y\in \myH$. This shows that $x_t$ is cointegrated, which is implication $(ii)$. Because $\myH$ is infinite dimensional, one has  $(\Ima C_0)^\bot$ is also infinite dimensional, which is implication $(iv)$.
\end{mproof}

\smallskip

\begin{mproof}{\textit{Proof of Proposition \tref{prop_ckp}}.}
First note that for $d=1$, Theorem \tref{thm_EXI} and Corollary \tref{coro_EXI_CI} imply $\Delta x_t=B(L)\varepsilon_t$, where $B(z)=\sum_{n=0}^{\infty} B_{n}z^n$ is absolutely convergent on $D(0,\rho)$, $\rho>1$, $B(1)=C_0\neq 0$ and $\Ima B(1)=\Ima C_0$ is finite dimensional. Because $B(z)$ is infinitely differentiable on $D(0,\rho)$, $\rho>1$, the series obtained by termwise differentiation coincides with the first derivative of $B(z)$ for each $z\in D(0,\rho)$, and hence one has $\sum_{n=1}^{\infty}n\|B_{n}\|_{\myB}<\infty$.
\end{mproof}

\smallskip

\begin{mproof}{\textit{Proof of Proposition \tref{prop_compact}}.}
Because the sum of compact operators is compact, see Theorem 16.1 in \citet[][Chapter II]{GGK:03}, and if $K$ is compact then $I-K$ is \fof, see Theorem 4.2 in \citet[][Chapter XV]{GGK:03}, then $A_0=I-\sum_{n=1}^kA^\circ_n$ is \fof. Because $A_0$ is non-invertible, by the Fredholm alternative there exist for nonzero $v\in \myH$ such that $A_0v=0$, see Theorem 4.1 in \citet[][Chapter XIII]{GGK:03}. Finally, since $z=1$ is assumed to be the only isolated singularity of $A(z)^{-1}$ within $D(0,\rho)$, $\rho>1$, this shows that $z=1$ is a eigenvalue of finite type of $A(z)$.
\end{mproof}

\smallskip

\begin{mproof}{\textit{Proof of Theorem \tref{thm_CHAR_1}}.}
Set $d=1$ in Theorem \tref{thm_CHAR_d}.
\end{mproof}

\smallskip

\begin{mproof}{\textit{Proof of Proposition \tref{prop_BSS_k_1}}.}
When $k=1$, \eqref{eq_BSS_I1} reads $\myH=\Ima A_0 \oplus \Ker A_0$. By assumption, $\Ker A_0$ has finite dimension; hence $\myH=\Ima A_0 \oplus \Ker A_0$ implies that $(\Ima A_0)^\bot$ has finite dimension equal to $\dim \Ker A_0$. This shows that $A_0$ is \fof. Because $\dim \Ker A_0 > 0$, there exist for nonzero $v\in \myH$ such that $A_0v=0$ and since $z=1$ is assumed to be the only isolated singularity of $A(z)^{-1}$ within $D(0,\rho)$, $\rho>1$, this shows that $z=1$ is a eigenvalue of finite type of $A(z)$.
\end{mproof}

\smallskip

\begin{mproof}{\textit{Proof of Proposition \tref{prop_equiv_BSS}}.}
The notation $\myalpha_0=\Ima A_0$ and $\mybeta_0=(\Ker A_0)^\bot$, see \eqref{eq_def_0}, is employed. In the present notation, \eqref{eq_BSS_I1} reads $\myH=\myalpha_0 \oplus A_1 \mybeta_0^\bot$, where by assumption of \ar, see Remark \tref{rem_eigFT0}, one has  $0<\dim \mybeta_0^\bot=\dim \myalpha_0^\bot<\infty$. Observe that \eqref{eq_I1_cond} is equivalent to $A_{0}C_{1}+A_{1}C_{0}= I$, see $(ii)$ in Theorem \tref{thm_pole_order} in Appendix \tref{app_lemmas}.

\noindent\eqref{eq_I1_cond} $\rimp$ \eqref{eq_BSS_I1}. Let \eqref{eq_I1_cond} hold, which is equivalent to
$A_{0}C_{1}+A_{1}C_{0} = I$ by Theorem \tref{thm_pole_order}. This implies that for any $v\in \myH$ one has $v=u+s$, where $u=A_{0}C_{1}v\in \myalpha_0$ because $\Ima A_{0} = \myalpha_0$ and $s=A_{1}C_{0}v\in A_1 \mybeta_0^\bot$ because $\Ima C_{0} = \mybeta_0^\bot$, see $(v)$ and $(vi)$ in Theorem \tref{thm_pole_order}.
In order to show that $\myalpha_0 \cap A_1 \mybeta_0^\bot = \{0\}$, note that $\Ker C_0=\myalpha_0$, see $(iii)$ and $(iv)$ in Theorem \tref{thm_pole_order}. This implies that for any $v\in \myalpha_0$ one has $s=A_{1}C_{0}v=0$, i.e. $s\neq 0 $ implies $P_{\myalpha_0^\bot}v\neq  0 $. However, $s\neq  0 $ belongs to $\myalpha_0$ if and only if $P_{\myalpha_0^\bot}s=P_{\myalpha_0^\bot}A_{1}C_{0}v=0$. Applying $P_{\myalpha_0^\bot}$ on both sides of $A_{0}C_{1}+A_{1}C_{0} = I$ one gets $P_{\myalpha_0^\bot}A_{1}C_{0} = P_{\myalpha_0^\bot}$; hence $P_{\myalpha_0^\bot}s=P_{\myalpha_0^\bot}A_{1}C_{0}v=P_{\myalpha_0^\bot}v\neq  0 $
which gives a contradiction. This shows that there does not exist $s\neq  0 $
that belongs to $\myalpha_0$, i.e. $\myalpha_0 \cap A_1 \mybeta_0^\bot=\{0\}$, so that \eqref{eq_BSS_I1} holds.

\noindent \eqref{eq_BSS_I1} $\rimp$ \eqref{eq_I1_cond}. Assume that $\myH=\myalpha_0 \oplus A_1 \mybeta_0^\bot$; because $A_{0}C_{1}+A_{1}C_{0} = 0$ implies that there exists a nonzero $v\in \myH$ such that $v=u+s=0$, where $u=A_{0}C_{1}v\in \myalpha_0$ and $s=A_{1}C_{0}v\in A_1 \mybeta_0^\bot$, this is a contradiction and hence \eqref{eq_I1_cond} holds.
\end{mproof}

\smallskip

\begin{mproof}{\textit{Proof of Theorem \tref{thm_CHAR_2}}.}
Set $d=2$ in Theorem \tref{thm_CHAR_d}.
\end{mproof}

\smallskip

\begin{mproof}{\textit{Proof of Theorem \tref{thm_CHAR_d}}.}
The proof makes use of Theorem \tref{thm_pole_order} in Appendix \tref{app_lemmas}, which establishes the order of integration of the process, and Theorem \tref{thm_pole_canc} in Appendix \tref{app_lemmas}, which describes the pole cancellations that give rise to cointegration. By Theorem \tref{thm_pole_order} one has  $A(z)^{-1}$ has a pole of order $d$ at $z=1$, i.e. $x_t\sim I(d)$, if and only if $\Ima C_0=\mybeta_d$, where $\mybeta_d=(\mybeta_0 \oplus \cdots \oplus \mybeta_{d-1})^\bot\neq \{ 0 \}$ . The common trends representation is found in \eqref{eq_CT}; because $\Ima C_0=\mybeta_d$ and $\mybeta_d=(\mybeta_0 \oplus \cdots \oplus \mybeta_{d-1})^\bot\neq \{ 0 \}$, this shows that $\langle v,x_t\rangle \sim I(d)$ for any nonzero $v\in \mybeta_d$, so that $\mybeta_d$ is the finite dimensional \as\ and $\mybeta_0 \oplus \cdots \oplus \mybeta_{d-1}$ is the infinite dimensional \cs. Finally, applying Theorem \tref{thm_pole_canc} one has that $\langle v,\gamma_{h}(z) A(z)^{-1}y\rangle$, where $\gamma_{h}(z)=P_{\mybeta_h}+S^+_h\sum_{n=1}^{d-h-1} A_{h+1,n}(1-z)^n$, has a pole of order $h=0,1,\dots,d$ for any nonzero $v\in \mybeta_h$ and any nonzero $y\in \myH$. This shows that $\langle v,x_t\rangle + \sum_{n=1}^{d-h-1}\langle v,S_h^+A_{h+1,n}\Delta^n x_t\rangle \sim I(h)$ for any nonzero $v\in \mybeta_h$, $h=0,1,\dots,d$.
\end{mproof}

\smallskip

\begin{mproof}{\textit{Proof of Proposition \tref{prop_equiv_HP}}.}
$A_P-I$ is a nilpotent matrix of order $d$ if and only if the largest Jordan block of $A_P$ with eigenvalue 1 has dimension $d$, see e.g. \citet[][p. 181]{HJ:13}. In section Section 4.4 of \citet{FP_idcoeff}, it is proved that the size of the largest Jordan block of a matrix is equal to $d$ if and only if the \pole{d} condition holds.
\end{mproof}

\end{document}